\documentclass[10pt, aps, superscriptaddress, nofootinbib ,floatfix, notitlepage]{revtex4-1}
\usepackage[utf8x]{inputenc}
\pdfoutput=1
\usepackage{graphicx}
\usepackage{hyperref}
\usepackage{bm}
\usepackage{slashed}
\usepackage{geometry}
\usepackage{graphicx}
\usepackage{amssymb}
\usepackage{epstopdf}
\usepackage{amsmath}
\usepackage{xcolor}
\usepackage{adjustbox}

\newcommand{\be}{\begin{equation}}
\newcommand{\ee}{\end{equation}}
\newcommand{\bea}{\begin{eqnarray}}
\newcommand{\eea}{\end{eqnarray}}

\newcommand{\LaSapienza}{Physics Department and INFN Sezione di Roma La Sapienza,\\ Piazzale Aldo Moro 5, 00185 Roma, Italy}
\newcommand{\RomatreINFN}{Istituto Nazionale di Fisica Nucleare, Sezione di Roma Tre,\\ Via della Vasca Navale 84, I-00146 Rome, Italy}
\newcommand{\Annecy}{LAPTh, Universit\'e Savoie Mont-Blanc and CNRS, F-74941 Annecy, France}

\begin{document}

\title{Semileptonic $B \to D^*$ decays from light to $\tau$ leptons:\\  the extraction of the form factor $F_2$ from data}

\author{G.\,Martinelli}\affiliation{\LaSapienza}
\author{S.\,Simula}\affiliation{\RomatreINFN}
\author{L.\,Vittorio}\affiliation{\Annecy}
 
\begin{abstract}
We extend the Standard Model (SM) analysis of Ref.\,\cite{Martinelli:2024vde}, which was limited to light leptons in the final state, to the  semileptonic $B \to D^* \tau \nu_\tau$  decay. By using quantities that can be analised without the knowledge of $\vert V_{cb}\vert$, we derive important information about the helicity amplitudes and the hadronic form factors that can be compared with the predictions of lattice QCD calculations. In  particular, there is a difficulty in reproducing simultaneously the experimental values  of $R(D^*)$ and of other quantities relevant for the semitauonic decays within the SM.
As a byproduct of our analysis, we also present a determination of $\vert V_{cb}\vert$ from the total decay rate.
\end{abstract}

\maketitle

\section{Introduction}
\label{sec:intro}
In this paper we extend the Standard Model (SM) analysis of Ref.\,\cite{Martinelli:2024vde}, which was limited to light leptons in the  final state,  to include quantities related to the  semileptonic $B \to D^* \tau \nu_\tau$  decay. In particular, we consider
\bea  
    && {\rm  the~ratio~of~the~branching~fractions}  \,\, R(D^*)= \frac{\Gamma(B \to D^* \tau \nu_\tau)}{\Gamma(B \to D^* \ell \nu_\ell)}\, , \nonumber \\[2mm] 
    \label{eq:inqua}  
    && {\rm  the~longitudinal~D}^*{\rm -polarization~fraction} \,\, F_{L, \tau}\, , \\ [3mm]
    && {\rm the}  \,\, \tau{\rm -polarization}\,\, P_\tau(D^*) \, . \nonumber
 \eea
In this way we may check the consistency of the measurements available for the quantities in Eq.\,(\ref {eq:inqua}) (see Eqs.\,(\ref{eq:RDstar_exp})-(\ref{eq:Ptau_exp}) below) within the SM. 

Using only the $\tau$-observables\,(\ref{eq:inqua}) and light-lepton quantities that can be computed without the knowledge of $\vert V_{cb}\vert$, we  may extract the {\em reduced} form factors (FFs) $g(w) / f(1)$, $f(w) / f(1)$, $F_1(w) / f(1)$ and $F_2(w) / f(1)$ from the data and compare them to the results of Lattice QCD (LQCD) calculations. The standard FFs $g(w)$, $f(w)$, $F_1(w)$ and $F_2(w)$ are those introduced in Ref.\,\cite{Boyd:1997kz}. They are expressed as functions of the recoil variable $w$, which in terms of the squared four-momentum transfer $q^2$  is given by
\be
    w \equiv \frac{1 + r^2 - q^2 / m_B^2}{2 r} ~ ,
    \label{eq:w}
\ee
with $r \equiv m_{D^*} / m_B$.

 In the light-lepton sector our SM analysis is based on the use of four single-differential ratios for semileptonic $B \to D^* \ell \nu_\ell$ decays (see Eqs.\,(\ref{eq:ratio_w})-(\ref{eq:ratio_chi}) below). Among them the angular distributions depend on three basic parameters (denoted in the following as $ \{ \eta, \delta, \epsilon\}$), which in the massless lepton limit are defined in terms of experimentally measurable quantities: the forward-backward asymmetry $A_{FB}$, the longitudinal $D^*$-polarization fraction $F_L$ and the transverse asymmetry $A_{1c}$ (see Eqs.\,(\ref{eq:AFB})-(\ref{eq:A1c}) below).

We show that the addition of the quantities in Eq.\,(\ref{eq:inqua}) to the analysis of the observables relevant for light-lepton semileptonic $B \to D^*$ decays leaves substantially unchanged the results of the fit of the reduced FFs $g(w) / f(1)$, $f(w) / f(1)$ and  $F_1(w) / f(1)$, while allowing for the extraction of the reduced FF $F_2(w) / f(1)$.  

The main results of our work are the following:
\begin{itemize}

\item The value of $F_2(w) / f(1)$ necessary to reproduce the experimental values of $F_{L, \tau}$, determined recently in two $w$-bins by the LHCb Collaboration\,\cite{LHCb:2023ssl}, is in good agreement with LQCD calculations. On the contrary, in order to fit the experimental determination of $R(D^*)$ a much larger value of $F_2(w) / f(1)$ is necessary in the whole relevant kinematical range.  
This difference may be due to some fluctuation/systematic effect in the experimental measurements or to the presence of New Physics (NP) affecting in a different way $F_{L, \tau}$ and $R(D^*)$ (see Refs.\,\cite{Blanke:2018yud, Blanke:2019qrx}). In any case the difference is within 2$\sigma$ deviations.

\item The large value of $F_2(w) / f(1)$, required by the addition of the experimental value of $R(D^*)$, yields values of other observables in semitauonic decays, namely $F_{L, \tau}$, $P_\tau(D^*)$, $A_{FB, \tau}$ and $A_{1c, \tau}$, in tension with the corresponding LQCD predictions up to $\approx 4\sigma$ level. In particular, we show that the SM implies a strong anti-correlation between $R(D^*)$ and $A_{FB, \tau}$, an observable not yet measured.

\item There is an apparent tension between the helicity amplitude $H_0$, obtained from the analysis of the light-lepton data of Refs.\,\cite{Belle:2018ezy, Belle:2023bwv, Belle-II:2023okj}, and the form factor $F_1$, used to compute $H_0$, obtained in Ref.\,\cite{Martinelli:2023fwm} using the Dispersive Matrix (DM) approach by combining the three different LQCD calculations from FNAL/MILC\,\cite{FermilabLattice:2021cdg}, HPQCD\,\cite{Harrison:2023dzh} and JLQCD\,\cite{Aoki:2023qpa}\footnote{The three lattice determinations of the FFs are uncorrelated, since the three Collaborations\,\cite{FermilabLattice:2021cdg, Harrison:2023dzh, Aoki:2023qpa} do not use common gauge ensembles.}. There is however a good consistency between $H_0$, obtained using only the data from Ref.\,\cite{Belle:2023bwv}, and $F_1$ (see also Ref.\,\cite{Martinelli:2024vde}). Finally, there is also a good consistency between $H_0$, obtained from the analysis of the light-lepton data of Refs.\,\cite{Belle:2018ezy, Belle:2023bwv, Belle-II:2023okj}, and $F_1$ from JLQCD only.
In what follows all the results corresponding to the combined FFs obtained in Ref.\,\cite{Martinelli:2023fwm} will be referred to as the LQCD predictions. The reader is also referred to Ref.\,\cite{Martinelli:2023fwm} for an extensive discussion about the differences among the FFs determined by each of the lattice Collaborations and their impact in the determination of $\vert V_{cb} \vert$ and $R(D^*)$.

\item The simultaneous addition of all the quantities given in Eq.\,(\ref{eq:inqua}) is dominated by the experimental value of $R(D^*)$ due to its current, higher precision (see Eqs.\,(\ref{eq:RDstar_exp})-(\ref{eq:Ptau_exp}) below).

\item Using the reduced FFs of our SM analysis of the light-lepton data of Refs.\,\cite{Belle:2018ezy, Belle:2023bwv, Belle-II:2023okj} and the total decay rate for the semileptonic $B \to D^* \ell \nu_\ell$ with $\ell = e, \mu$ from the latest PDG review\,\cite{ParticleDataGroup:2024cfk}, we obtain $|V_{cb} f(1)| = 0.2318 (46)$\,GeV. Assuming for the FF value $f(1)$ the LQCD prediction $f(1) = 5.845(50)$\,GeV (see Eq.\,(\ref{eq:f1}) below), this implies $|V_{cb}| = 39.7 (8) \cdot 10^{-3}$. This value agrees well with the result $|V_{cb}| = 39.92 (64) \cdot 10^{-3}$, obtained by the bin-per-bin analysis of Ref.\,\cite{Martinelli:2023fwm} carried out using the same experimental data sets and lattice points. It agrees also with the result $|V_{cb}| = 39.46 (53) \cdot 10^{-3}$ from the FLAG review\,\cite{FlavourLatticeAveragingGroupFLAG:2024oxs}, while it deviates by $\approx 2.6 \sigma$ from the latest determination $|V_{cb}| = 42.00 (47) \cdot 10^{-3}$\,\cite{Fael:2024rys} from inclusive semileptonic decays and  by $\approx 2.7 \sigma$ from the updated prediction $|V_{cb}| = 42.22 (51) \cdot 10^{-3}$ obtained within the SM by UTfit\,\cite{UTfit:2022hsi}.

\end{itemize}

The plan of the remainder of the paper is the following: in Sec.\,\ref{sec:rate} we  recall the most general expression of the $B \to D^*$ differential decay rate in the recoil variable and in the relevant angular variables; in Sec.\,\ref{sec:ratios} we recall the basic formulae  for the partially integrated differential rates and introduce the  basic parameters $ \{ \eta,  \delta, \epsilon\}$  in terms of the helicity amplitudes computed in the SM (and in the massless lepton limit);  in Sec.\,\ref{sec:data} we extend the formulae used to analyse the data, given in Ref.\,\cite{Martinelli:2024vde} for light leptons, to the case  $B \to D^* \tau \nu_\tau $ and list the experimental values and uncertainties of the quantities in Eq.\,(\ref{eq:inqua}) used in this work; in Sec.\,\ref{sec:BGL} we recall the $z$-expansion approach developed in Ref.\,\cite{Boyd:1997kz}, commonly referred to as the Grinstein-Boyd-Lebed (BGL) expansion, which has been used in our analysis to fit the FFs to the experiments data; in Secs.\, \ref{sec:light} and \, \ref{sec:light+tau} we describe the fit of the quantities related to light-lepton or combined light lepton + $\tau$ semileptonic decays; in Sec.\,\ref{sec:Vcbf1}, although this is not the main subject of our paper, we also determine the CKM matrix element $\vert V_{cb}\vert$ using the total decay rate $\Gamma(B \to D^* \ell \nu_\ell)$ for light leptons.
 The final Section contains our  conclusions  and some ancillary material is given in the appendices.

\section{The $B \to D^*$ differential decay rate for massless leptons}
\label{sec:rate}

\begin{figure}[htb!]
\begin{center}
\includegraphics[scale=1.25]{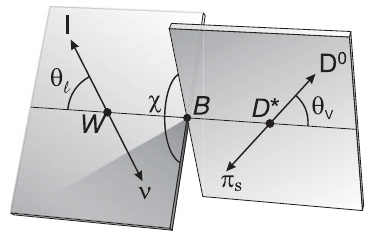}
\end{center}
\vspace{-0.5cm}
\caption{\it \small Definition of the angles $\theta_v$, $\theta_\ell$ and $\chi$ for the decay $B \to D^*(D \pi) \ell \nu_\ell$ (from Ref.\,\cite{Belle:2018ezy}).}
\label{fig:angles}
\end{figure}

Using the helicity basis for the hadronic form factors (FFs), the four-fold differential rate for the semileptonic $B \to D^* \ell \nu_\ell$ decay, within the Standard Model (SM) and  in the massless lepton limit ($m_\ell = 0$), can be written as
\bea
    \frac{d^4 \Gamma}{dw d\mbox{cos}\theta_v d\mbox{cos}\theta_\ell d\chi} & = & \frac{3}{16\pi} \Gamma_0 \sqrt{w^2 - 1} (1 - 2r w + r^2) 
                  \Big\{ H_+^2(w) ~ \mbox{sin}^2\theta_v ~ (1 - \mbox{cos}\theta_\ell)^2 \nonumber \\[2mm]
        & + & H_-^2(w) ~ \mbox{sin}^2\theta_v ~ (1 + \mbox{cos}\theta_\ell)^2 + 4 ~ H_0^2(w) ~ 
                  \mbox{cos}^2\theta_v ~ \mbox{sin}^2\theta_\ell \nonumber \\[2mm]
        & - & 2 ~ H_-(w) H_+(w) ~ \mbox{sin}^2\theta_v ~ \mbox{sin}^2\theta_\ell  ~ \mbox{cos}2\chi \nonumber \\[2mm]
        & - &  2 ~ H_+(w) H_0(w) ~ \mbox{sin}2\theta_v ~ \mbox{sin}\theta_\ell  ~ (1 - \mbox{cos}\theta_\ell) ~ \mbox{cos}\chi \nonumber \\ [2mm]
        & + &  2 ~ H_-(w) H_0(w) ~ \mbox{sin}2\theta_v ~ \mbox{sin}\theta_\ell  ~ (1 + \mbox{cos}\theta_\ell) ~ \mbox{cos}\chi \Big\} ~ , ~
        \label{eq:Gamma4_ell}
\eea
where
\be
    \Gamma_0 \equiv \frac{\eta_{EW}^2 m_B m_{D^*}^2}{(4 \pi)^3} G_F^2 |V_{cb}|^2 
    \label{eq:Gamma_0}
\ee
with $\eta_{EW} = 1 + \alpha \ln(M_Z/m_B)/\pi \simeq 1.0066$ the leading electromagnetic correction, $G_F$ the Fermi constant and $V_{cb}$ the relevant CKM matrix element.
The angles $\theta_v$, $\theta_\ell$ and $\chi$ are defined as in Fig.~\ref{fig:angles}.

The three helicity FFs $H_{+,-,0}(w)$ are related to the standard FFs $g(w)$, $f(w)$, $F_1(w)$ of Ref.\,\cite{Boyd:1997kz}, corresponding to definite spin-parity (to which the unitarity bounds can be applied), by
\bea
    \label{eq:H+}
    H_+(w) & = & f(w) - m_B^2 r \sqrt{w^2 -1} ~ g(w) ~ , ~ \\[2mm]
    \label{eq:H-}
    H_-(w) & = & f(w) + m_B^2 r \sqrt{w^2 -1} ~ g(w) ~ , ~ \\[2mm]
    \label{eq:H0}
    H_0(w) & = & \frac{F_1(w) }{m_B \sqrt{1 - 2 r w + r^2}}  ~  . ~
\eea

The fourth (``time-like") helicity amplitude $H_t(w)$ contributes to the differential decay rate only for heavy leptons and it is related to the pseudoscalar form factor $F_2(w)$ by
\be
     \label{eq:Ht}
    H_t(w) = \frac{m_B r \sqrt{w^2 - 1}}{\sqrt{1 - 2 r w + r^2}} ~ F_2(w) ~  . ~
\ee

The hadronic FFs satisfy exactly two kinematical constraints (KCs) at $w = 1$ and $w = w_{max} \equiv (1 + r^2) / (2r)$:
\bea
      \label{eq:KC1}
      F_1(w=1) & = & m_B (1 - r) ~ f(w=1) ~ , ~ \\[2mm]
      \label{eq:KC2}
      F_2(w=w_{max}) & = & \frac{2}{m_B^2 (1 - r^2)} ~ F_1(w=w_{max}) ~ . ~
\eea
The above KCs are specific of the FFs with definite spin-parity. We remind that the KC at $w = 1$ comes from the fact that at zero recoil the three helicity amplitudes for the final $D^*$-meson at rest are not independent. The second KC at $w = w_{max}$ is related to the cancellation of any apparent kinematical singularity at $q^2 = 0$ in the Lorentz decomposition of the matrix elements of the $(V - A)$ weak current.

After integration over the recoil variable $w$ (from $1$ to $w_{max}$), the angular variables $\theta_v$ (from $0$ to $\pi$),  $\theta_\ell$ (from $0$ to $\pi$) and $\chi$ (from $0$ to $2 \pi$), the total decay rate for massless leptons is given by
\be
    \label{eq:Gamma_ell}
    \Gamma \equiv \Gamma(B \to D^* \ell \nu_\ell)|_{m_\ell = 0} = \frac{4}{3} \Gamma_0  \overline{\cal{H}} ~ , ~
\ee
where 
\be
     \label{eq:H}
      \overline{\cal{H}} \equiv \overline{H}_{++} + \overline{H}_{--} + \overline{H}_{00} ~ , ~ 
\ee
and $\overline{H}_{++, --, 00}$ are the diagonal elements of the matrix
\be
    \label{eq:Hij}
    \overline{H}_{ij} \equiv \int_1^{w_{max}} dw \sqrt{w^2 - 1} (1 - 2r w + r^2) ~  H_i(w) ~ H_j(w) ~ , ~
\ee
that is defined in terms of  six independent elements.

The above matrix elements determine completely the hadronic contribution to the angular structure of the ratio between the triple-differential angular and the total decay rate, namely
\bea
        \frac{1}{\Gamma} \frac{d^3 \Gamma}{d\mbox{cos}\theta_v d\mbox{cos}\theta_\ell d\chi} & = &  \frac{9}{64 \pi} 
                \frac{1}{\overline{\cal{H}}} \Big\{ \overline{H}_{++} ~ \mbox{sin}^2\theta_v ~ (1 - \mbox{cos}\theta_\ell)^2 + 
               \overline{H}_{--} ~ \mbox{sin}^2\theta_v ~ (1 + \mbox{cos}\theta_\ell)^2 ~ \qquad \nonumber \\[2mm]
        & + &  4 ~ \overline{H}_{00} ~ \mbox{cos}^2\theta_v ~ \mbox{sin}^2\theta_\ell - 
                  2 ~ \overline{H}_{+-} ~ \mbox{sin}^2\theta_v ~ \mbox{sin}^2\theta_\ell  ~ \mbox{cos}2\chi \nonumber \\[2mm]
        & - &  2 ~ \overline{H}_{0+} ~ \mbox{sin}2\theta_v ~ \mbox{sin}\theta_\ell  ~ (1 - \mbox{cos}\theta_\ell) ~ \mbox{cos}\chi \nonumber \\ [2mm]
        & + &  2 ~ \overline{H}_{0-} ~ \mbox{sin}2\theta_v ~ \mbox{sin}\theta_\ell  ~ (1 + \mbox{cos}\theta_\ell) ~ \mbox{cos}\chi \Big\} ~ . ~
        \label{eq:Gamma3_ell}
\eea
Note that the CKM element $|V_{cb}|$ does not contribute to Eq.\,(\ref{eq:Gamma3_ell}) and, therefore, any analysis based on Eq.\,(\ref{eq:Gamma3_ell}) never requires the knowledge of  $|V_{cb}|$. The only assumptions are the SM and the massless lepton limit\footnote{We have explicitly checked that the inclusion of muon mass effects are negligible in our SM analysis.}.

\section{The partially integrated differential decay rates}
\label{sec:ratios}

Starting from Eq.\,(\ref{eq:Gamma4_ell}) one can construct four partially integrated differential decay rates $d\Gamma / dx$ where $x = \{ w, \mbox{cos}(\theta_v), \mbox{cos}(\theta_{\ell}), \chi \}$ and divide them by the total decay rate (\ref{eq:Gamma_ell}).
One gets\,\cite{Bobeth:2021lya}
\bea
        \label{eq:ratio_w}
        \frac{1}{\Gamma} \frac{d\Gamma}{dw} & = & \sqrt{w^2 - 1} (1 - 2r w + r^2)  ~ \frac{H_+^2(w) + H_-^2(w) + H_0^2(w)}{\overline{\cal{H}}} ~ , ~ \\[2mm]
        \label{eq:ratio_v}
        \frac{1}{\Gamma} \frac{d\Gamma}{d\mbox{cos}(\theta_v)} & = & \frac{3}{4(1+ \eta) } ~ \left[ \eta+ (2 - \eta) ~ \mbox{cos}^2(\theta_v) \right] ~ , ~ \\[2mm]
         \label{eq:ratio_ell}
        \frac{1}{\Gamma} \frac{d\Gamma}{d\mbox{cos}(\theta_\ell)} & = & \frac{3}{8(1+ \eta)} ~ \left[ 2 + \eta - 2 \delta ~ \mbox{cos}(\theta_\ell) -
                                                                                                                 (2 - \eta) ~ \mbox{cos}^2(\theta_\ell) \right] ~ , ~ \\[2mm]
        \label{eq:ratio_chi}
        \frac{1}{\Gamma} \frac{d\Gamma}{d\chi} & = & \frac{1}{2 \pi} \left[ 1 - \frac{\epsilon}{1 + \eta} ~ \mbox{cos}(2\chi) \right] ~ , ~
\eea
where the hadronic quantities $\eta, \delta, \epsilon$ (already introduced in Ref.\,\cite{Martinelli:2024vde}) are given explicitly by
\bea
        \label{eq:eta}
        \eta & = & \frac{\overline{H}_{++} + \overline{H}_{--}}{\overline{H}_{00}} ~ , ~ \\[2mm]
        \label{eq:delta}
        \delta & = & \frac{\overline{H}_{++} - \overline{H}_{--}}{\overline{H}_{00}} ~ , ~ \\[2mm]
         \label{eq:epsilon}
        \epsilon & = & \frac{\overline{H}_{+-}}{\overline{H}_{00}} ~ . ~ 
\eea
The standard forward-backward asymmetry $A_{FB}$,  the longitudinal $D^*$-polarization fraction $F_L$ and  the transverse asymmetry $A_{1c}$ can be expressed in terms of the quantities defined in Eqs.\,(\ref{eq:eta})-(\ref{eq:epsilon}) as\,\cite{Martinelli:2024vde}
\bea
     \label{eq:AFB}
     A_{FB} & = & - \frac{3}{4} \frac{\delta}{1 + \eta} = \frac{3}{4} \frac{\overline{H}_{--} - \overline{H}_{++}}{\overline{\cal{H}}} ~ , ~ \\[2mm]
      \label{eq:FL}
     F_L & = & \frac{1}{1 + \eta} = \frac{\overline{H}_{00}}{\overline{\cal{H}}}  ~ , ~ \\[2mm]
     \label{eq:A1c}
     A_{1c} & = & - \frac{\epsilon}{1 + \eta} = - \frac{\overline{H}_{+-}}{\overline{\cal{H}}} ~ . ~
\eea
For analogy, taking into account the angular structure of Eq.\,(\ref{eq:Gamma3_ell}), two other transverse asymmetries can be defined
\bea
     \label{eq:A2c}
     A_{2c} & = &  - \frac{\overline{H}_{0+}}{\overline{\cal{H}}}~ , ~ \\[2mm]
     \label{eq:A3c}
     A_{3c} & = & - \frac{\overline{H}_{0-}}{\overline{\cal{H}}} ~ , ~
\eea
which however do not appear in Eqs.\,(\ref{eq:ratio_w})-(\ref{eq:ratio_chi}). Their determination requires the analysis of the triple differential ratio\,(\ref{eq:Gamma3_ell}).

We stress that from Eqs.\,(\ref{eq:ratio_v})-(\ref{eq:ratio_chi}), which hold in the SM and in the massless lepton limit, the three angular observables $(1 / \Gamma) d\Gamma / dx$ with $x = \{\mbox{cos}(\theta_v), \mbox{cos}(\theta_{\ell}), \chi \}$ depend only on three hadronic quantities $\{\eta, \delta, \epsilon\}$, given in Eqs.\,(\ref{eq:eta})-(\ref{eq:epsilon}), or, equivalently, on $A_{FB}$, $F_L$ and $A_{1c}$, given by Eqs.\,(\ref{eq:AFB})-(\ref{eq:A1c}).

\section{Experimental data}
\label{sec:data}

In the light-lepton sector, following Ref.\,\cite{Martinelli:2024vde}, we consider the three experimental data sets available for the four single-differential decay rates $d\Gamma / dx$, where $x = \{ w, \mbox{cos}(\theta_v), \mbox{cos}(\theta_{\ell}), \chi \}$, from Refs.\,\cite{Belle:2018ezy, Belle:2023bwv, Belle-II:2023okj}, which hereafter will be labelled as Belle18, Belle23 and BelleII23, respectively\footnote{More precisely, the Belle18 data set is provided in terms of the quantities $d\Gamma / dx$, while both Belle23 and BelleII23 data sets are available directly in terms of the ratios $(1/\Gamma) d\Gamma / dx$.}.

For each of the kinematical variables we have $N_x$ experimental bins, so that we consider the ratios
\be
     \label{eq:ratios}
     R_n(x) \equiv \frac{1}{\sum_{m = 1}^{N_x} \Delta \Gamma_m(x)} \Delta \Gamma_n(x) ~ , \quad n = 1, 2, ... N_x ~ , ~
\ee
where 
\be
      \Delta \Gamma_n(x) = \int_{x_{n-1}}^{x_n} dx \frac{d\Gamma}{dx}  ~ . ~
      \label{eq:Gamma_bins}
\ee
We stress that the ratios\,(\ref{eq:ratios}) are independent of $|V_{cb}|$ and satisfy the normalization condition $\sum_{n = 1}^{N_x} R_n(x) = 1$.
The details of the experimental bins for the Belle18, Belle23 and BelleII23 data sets are summarised in Appendix\,\ref{sec:bins}, where also explicit formulae for the ratios\,(\ref{eq:ratios}) are provided.

 It is  clear that from the quantities considered in the previous section one can determine the hadronic FFs only up to a common multiplicative constant. We choose such constant to be the value of the form factor $f(w)$ at zero recoil, i.e.~$f(1) = f(w=1)$.  
In addition, in order to extract from the experiments  all the four hadronic FFs $g$, $f$, $F_1$ and $F_2$ it is necessary to add to the ratios\,(\ref{eq:ratios}) other quantities, independent of $|V_{cb}|$, and related to the case of heavy $\tau$-lepton.

In this work we consider the following $\tau$-observables:
\begin{itemize}

\item the branching ratio $R(D^*)$
\bea
     \label{eq:RDstar}
     R(D^*) & = & \frac{\Gamma(B \to D^* \tau \nu_\tau)}{\Gamma(B \to D^* \ell \nu_\ell)}  =  \frac{ \overline{\cal{H}}^\tau +  \frac{m_\tau^2}{2 m_B^2} \left(  \overline{\cal{L}}^\tau + 
                          3 \overline{L}_{tt}^\tau \right)}{ \overline{\cal{H}}} ~ , ~
\eea
where 
\bea
    \label{eq:Htau}
    \overline {\cal{H}}^\tau & \equiv & \overline{H}_{++}^\tau+ \overline{H}_{--}^\tau + \overline{H}_{00}^\tau ~ , ~ \\[2mm]
    \label{eq:Ltau}
      \overline{\cal{L}}^\tau & \equiv & \overline{L}_{++}^\tau + \overline{L}_{--}^\tau + \overline{L}_{00}^\tau ~ , ~ \\[2mm]
    \label{eq:Hii_tau}
    \overline{H}_{++,--,00}^\tau & = & \int_1^{w_{max}^\tau} dw \sqrt{w^2 - 1} ~ \left( 1 + r^2 - 2 r w \right) \left( 1 - \frac{m_\tau^2}{q^2} \right)^2 
                                                         H_{+,-,0}^2(w) ~ , ~ \quad \\[2mm]
    \label{eq:Lij_tau}                                                         
    \overline{L}_{++,--,00,tt}^\tau & = & \int_1^{w_{max}^\tau} dw \sqrt{w^2 - 1}  \left( 1 - \frac{m_\tau^2}{q^2} \right)^2 H_{+,-,0,t}^2(w) ~ 
\eea
with $w_{max}^\tau = \left( 1 + r^2 - \frac{m_\tau^2}{m_B^2} \right) / (2r)$;

\item the longitudinal $D^*$-polarization fraction $F_{L, \tau}$
\be
     \label{eq:FL_tau}
     F_{L, \tau} = \frac{\overline{H}_{00}^\tau + \frac{m_\tau^2}{2 m_B^2} \left( \overline{L}_{00}^\tau + 3 \overline{L}_{tt}^\tau \right)}
                         { \overline{\cal{H}}^\tau +  \frac{m_\tau^2}{2 m_B^2} \left(  \overline{\cal{L}}^\tau + 3 \overline{L}_{tt}^\tau \right)} ~ ; ~
\ee

\item the $\tau$-polarization $P_\tau(D^*)$
\be
     \label{eq:PDstar_tau}
     P_\tau(D^*) = 1 - \frac{2  \overline{\cal{H}}^\tau}{ \overline{\cal{H}}^\tau +  \frac{m_\tau^2}{2 m_B^2} \left( \overline{\cal{L}}^\tau + 3 \overline{L}_{tt}^\tau \right)}  ~ . ~
\ee

\end{itemize}

For the branching ratio $R(D^*)$ we use the latest experimental world average from Ref.\,\cite{Banerjee:2024znd}
\be
     \label{eq:RDstar_exp}
     R(D^*) = 0.286 \pm 0.012 ~ . ~
\ee

As for $F_{L, \tau}$, following Ref.\,\cite{Martinelli:2023fwm}, we average the experimental results $F_{L, \tau} = 0.60 (8) (4)$ from Ref.\,\cite{Belle:2019ewo} and $F_{L, \tau} = 0.41 (6) (3)$ from Ref.\,\cite{LHCb:2023ssl}, obtaining 
\be
    \label{eq:FLtau_exp}
    F_{L, \tau} = 0.48 \pm 0.09 ~ . ~
\ee
Furthermore, the longitudinal $D^*$-polarization fraction $F_{L, \tau}$ has been measured by LHCb\,\cite{LHCb:2023ssl} in two different $q^2$-bins: $q^2 < 7$ GeV$^2$ (low-$q^2$) and $q^2 > 7$ GeV$^2$ (high-$q^2$), namely
\bea
     \label{eq:FLtau_low}
     (F_{L, \tau})_< & = & 0.52 \pm 0.08 ~ \qquad q^2 < 7\,\mbox{GeV}^2 ~ , ~ \\[2mm]
     \label{eq:FLtau_high}
     (F_{L, \tau})_> & = & 0.34 \pm 0.08 ~ \qquad q^2 > 7\,\mbox{GeV}^2 ~
\eea
with a correlation coefficient equal to $-0.18$.
The explicit formulae for these two quantities are similar to Eqs.\,(\ref{eq:FL_tau}) and\,(\ref{eq:eta_SM}) with the elements $\overline{H}_{++, --, 00}^\tau$ and $\overline{L}_{++, --, 00,tt}^\tau$ replaced by integrals limited to the corresponding $q^2$-bins.

For $P_\tau(D^*)$ we consider the experimental value from Ref.\,\cite{Belle:2016dyj}
\be
    \label{eq:Ptau_exp}
    P_\tau(D^*) = -0.38 \pm 0.51^{+0.21}_{-0.16} = -0.36 \pm 0.54 ~ . ~
\ee

\section{The $z$-expansion of the hadronic form factors}
\label{sec:BGL}

In order to extract the hadronic FFs from the experimental data set we make use of the $z$-expansion approach developed in Ref.\,\cite{Boyd:1997kz}, commonly referred to as the Grinstein-Boyd-Lebed (BGL) expansion\footnote{In this work we do not use the DM method of Refs.\,\cite{Martinelli:2021myh, Martinelli:2023fwm}, since it would require a dedicated analysis of double differential distributions of the form $d^2\Gamma / dw dx$ with $x = \{\mbox{cos}(\theta_v), \mbox{cos}(\theta_{\ell}), \chi \}$. This is out of the scope of the present work.}.

Using the conformal variable $z$, defined as
\be
    \label{eq:z}
    z = \frac{\sqrt{t_+ - q^2} - \sqrt{t_+ - t_-}}{\sqrt{t_+ - q^2} + \sqrt{t_+ - t_-}} = \frac{\sqrt{1 + w} - \sqrt{2}}{\sqrt{1 + w} + \sqrt{2}}
\ee
with $t_\pm \equiv (m_B \pm m_{D^*})^2$, the hadronic FFs with definite spin-parity can be written as 
\bea
      \label{eq:g_BGL}
      g(w) & = & \frac{1}{\phi_g(z) B_g(z)} \sum_{i = 0}^\infty a_i^g z^i ~ , ~ \nonumber \\[2mm]
     \label{eq:f_BGL}
      f(w) & = & \frac{1}{\phi_f(z) B_f(z)} \sum_{i = 0}^\infty a_i^f z^i ~ , ~ \\[2mm]
     \label{eq:F1_BGL}
      F_1(w) & = & \frac{1}{\phi_{F_1}(z) B_{F_1}(z)} \sum_{i = 0}^\infty a_i^{F_1} z^i ~ , ~ \nonumber \\[2mm]
     \label{eq:F2_BGL}
      F_2(w) & = & \frac{1}{\phi_{F_2}(z) B_{F_2}(z)} \sum_{i = 0}^\infty a_i^{F_2} z^i ~ , ~ \nonumber 
\eea
where $a_i^{g, f, F_1, F_2}$ are real parameters and the kinematical functions $\phi_{g, f, F_1, F_2}(z)$, explicitly given by
\bea
    \label{eq:phig}
    \phi_{g}(z) & = & 16 r^2 \sqrt{\frac{n_I}{3\pi \chi_{1^-}}} \frac{(1 + z)^{2}}{\sqrt{1 - z}\left[ (1 + r)(1 - z) + 2 \sqrt{r}(1 + z) \right]^4} ~ , ~ \nonumber \\[2mm] 
    \label{eq:phif}
    \phi_{f}(z) & = & 4 \frac{r}{m_B^2} \sqrt{\frac{n_I}{3\pi \chi_{1^+}}} \, \frac{(1 + z)(1 - z)^{3/2}}{\left[ (1 + r)(1 - z) + 2 \sqrt{r}(1 + z) \right]^4} ~ , ~ \\ [2mm] 
    \label{eq:phiF1}
    \phi_{F_1}(z) & = & 4 \frac{r}{m_B^3} \sqrt{\frac{n_I}{6\pi \chi_{1^+}}} \frac{(1 + z)(1 - z)^{5/2}}{\left[ (1 + r)(1 - z) + 2 \sqrt{r}(1 + z) \right]^5} ~ , ~ \nonumber \\[2mm] 
    \label{eq:phiF2}
    \phi_{F_2}(z) & = &  16 r^2 \sqrt{\frac{n_I}{2\pi \chi_{0^-}}} \frac{(1 + z)^{2}}{\sqrt{1 - z}\left[ (1 + r)(1 - z) + 2 \sqrt{r}(1 + z) \right]^4} ~  , ~\nonumber 
\eea
include the values of the susceptibilities $\chi_{1^-, 1^+, 0^-}$ (derivatives of the vacuum polarization functions) relevant for the given spin-parity channels of the $B \to D^* \ell \nu_\ell$ decays\footnote{In Eqs.\,(\ref{eq:phif}) $n_I$ is an isospin Clebsh-Gordan factor. Following Refs.\,\cite{Martinelli:2021onb, Martinelli:2021myh, Martinelli:2023fwm} we adopt the value $n_I = 2$, which for the unitarity constraints represents a more conservative choice with respect to, e.g., Ref.\,\cite{Bigi:2017njr}.}. 
In Eqs.\,(\ref{eq:f_BGL}) the factors $B_{g, f, F_1, F_2}(z)$ take into account the presence of resonances below the pair production threshold\,\cite{Lellouch:1995yv}
\be
    \label{eq:poles}
    B(z) = \prod_R \frac{z - z(m_R^2)}{1 - z \, z(m_R^2)} ~ , ~
\ee
where $m_R$ is the mass of the resonance $R$ for each specific channel.
For the masses of the various poles corresponding to $B_c^{(*)}$ mesons we refer to Table III of Ref.~\cite{Bigi:2017jbd}.
We note that in the case of the $B \to D^*$ decay the conformal variable $z$ (see Eq.\,(\ref{eq:z})) ranges from $z(q^2 = t_-) = 0$ to $z(q^2 = 0)  =z_{max} \simeq 0.056$, while $z(m_R^2) < 0$. Thus,  the accurate location of the poles in Eq.\,(\ref{eq:poles}) is not crucial.

According to Ref.\,\cite{Boyd:1997kz} there are three unitarity constraints on the  BGL expansions\,(\ref{eq:f_BGL}), namely 
\bea
   \label{eq:UTfilter1}
   \sum_{i = 0}^\infty (a_i^g)^2 & \leq & 1 ~ , ~ \nonumber \\[2mm]
   \label{eq:UTfilter2}
    \sum_{i = 0}^\infty \left[ (a_i^f)^2 + (a_i^{F_1})^2 \right] & \leq & 1~ , ~ \\[2mm]
   \label{eq:UTfilter3}
   \sum_{i = 0}^\infty (a_i^{F_2})^2 & \leq & 1 ~ . ~ \nonumber
\eea

In  the analysis of the experimental data on the ratios $R_n(x)$ and on the $\tau$-observables we employ a truncated version of the $z$-expansions\,(\ref{eq:f_BGL}), including monomials $z^i$ up to given orders, which will be denoted by ($N_g$, $N_f$, $N_{F_1}$, $N_{F_2}$).

As already pointed out in the previous Section, our analysis can determine the hadronic FFs only up to a common multiplicative constant, chosen to be the value of the form factor $f(w)$ at zero recoil, i.e.~$f(1) = f(w=1)$.
Therefore, we introduce the ``reduced" FFs $\widetilde{g}$, $\widetilde{f}$, $\widetilde{F}_1$ and $\widetilde{F}_2$, defined as 
\bea
      \label{eq:gbar_BGL}
      \widetilde{g}(w) & = & \frac{g(w)}{f(1)} = \frac{1}{\widetilde{\phi}_g(z) B_g(z)} \sum_{i = 0}^{N_g} a_i^g z^i ~ , ~ \nonumber \\[2mm]
      \label{eq:fbar_BGL}
      \widetilde{f}(w) & = & \frac{f(w)}{f(1)} = \frac{1}{\widetilde{\phi}_f(z) B_f(z)} \sum_{i = 0}^{N_f} a_i^f z^i ~ , ~ \\[2mm]
      \label{eq:F1bar_BGL}
      \widetilde{F_1}(w) & = & \frac{F_1(w)}{f(1)} = \frac{1}{\widetilde{\phi}_{F_1}(z) B_{F_1}(z)} \sum_{i = 0}^{N_{F_1}} a_i^{F_1} z^i ~ , ~ \nonumber \\[2mm]
      \label{eq:F2bar_BGL}
      \widetilde{F_2}(w) & = & \frac{F_2(w)}{f(1)} = \frac{1}{\widetilde{\phi}_{F_2}(z) B_{F_2}(z)} \sum_{i = 0}^{N_{F_2}} a_i^{F_2} z^i ~ , ~ \nonumber 
\eea
where the kinematical functions $\widetilde{\phi}_{g, f, F_1, F_2}(z)$ are given by Eqs.\,(\ref{eq:phif}) with the susceptibilities $\chi_{1^-, 1^+, 0^-}$ replaced by $\chi_{1^-, 1^+, 0^-} / f^2(1)$. Moreover, the coefficient $a_0^f$ is given by
\be
     \label{eq:a0f}
     a_0^f = \widetilde{\phi}_f(0) B_f(0) = \frac{4r}{m_B^2} B_f(0) \sqrt{\frac{n_I}{3\pi}  \frac{f^2(1)}{\chi_{1^+}}} \frac{1}{(1 + \sqrt{r})^8} ~ , ~
\ee
while the coefficients $a_0^{F_1}$ and $a_0^{F_2}$ are fixed by the two KCs (\ref{eq:KC1}) and (\ref{eq:KC2}) in a straightforward way. Thus, the total number of free parameters in the $z$-expansions\,(\ref{eq:fbar_BGL}) is $N_{parms} = N_g + N_f + N_{F_1} + N_{F_2} + 1$.

Summarizing, the implementation of unitarity in the analysis of the experimental data on the ratios $R_n(x)$ and on the $\tau$-observables require the knowledge of only three theoretical inputs: the ``reduced" susceptibilities $\chi_{1^-, 1^+, 0^-} / f^2(1)$.
The nonperturbative values of the susceptibilities relevant for $b \to c$ weak transitions have been computed on the lattice in Ref.\,\cite{Martinelli:2021frl} (see also the recent determinations from Ref.\,\cite{Harrison:2024iad}\footnote{The results of Ref.\,\cite{Harrison:2024iad} for the total susceptibilities are compatible with the corresponding ones of Ref.\,\cite{Martinelli:2021frl} at $1.2 \sigma$, $0.4 \sigma$ and $1.7 \sigma$ in the spin-parity channels $0^-$, $1^-$ and $1^+$ relevant in this work. We have checked that such differences have a negligible impact on the widths of the bands of the reduced FFs extracted from the experimental data (see below Fig.\,\ref{fig:ffs_noPol}).}). 
After subtraction of the contribution of bound states one gets\,\cite{Martinelli:2021frl}
\bea
      \label{eq:bound1-}
      \chi_{1^-} & = & (5.84 \pm 0.44) \cdot 10^{-4} ~ \mbox{GeV}^{-2} ~ , ~ \nonumber \\[2mm]
      \label{eq:bound1+}
      \chi_{1^+} & = & (4.69 \pm 0.30) \cdot 10^{-4} ~ \mbox{GeV}^{-2} ~ ,~ \\[2mm]
      \label{eq:bound0-}
      \chi_{0^-} & = & (21.9 \pm 1.9) \cdot 10^{-3} ~ . ~\nonumber
\eea
Concerning $f(1)$ we adopt the value
\be
    \label{eq:f1}
    f(1) = 5.845 \pm 0.050 ~ \mbox{GeV} ~ , ~
 \ee
obtained in Ref.\,\cite{Martinelli:2023fwm} by the dispersive matrix (DM) method using as inputs all available lattice QCD (LQCD) results for the hadronic FFs, namely from FNAL/MILC\,\cite{FermilabLattice:2021cdg}, HPQCD\,\cite{Harrison:2023dzh} and JLQCD\,\cite{Aoki:2023qpa} Collaborations.

We close this Section by stressing that the theoretical input\,(\ref{eq:f1}) is used only in combination with the susceptibilities\,(\ref{eq:bound1+}) to impose the appropriate unitarity bounds on the reduced FFs\,(\ref{eq:fbar_BGL}).

\section{Analysis in the light-lepton sector}
\label{sec:light}

We begin our analysis starting from the light-lepton sector by considering only the experimental Belle18\,\cite{Belle:2018ezy}, Belle23\,\cite{ Belle:2023bwv} and BelleII23\,\cite{Belle-II:2023okj} data sets on the ratios $R_n(x)$ (see Eq.\,(\ref{eq:ratios}) with $x = \{ w, \mbox{cos}(\theta_v), \mbox{cos}(\theta_{\ell}), \chi \}$. 

From these data one can extract three out of the four reduced FFs, given by Eqs.\,(\ref{eq:fbar_BGL}), namely $\widetilde{g}$, $\widetilde{f}$ and $\widetilde{F}_1$, while for the FF $\widetilde{F}_2$ we include, for the time being, only the constant term $a_0^{F_2}$ fixed by the KC\,in Eq.\,(\ref{eq:KC2}) at $q^2 = 0$  (i.e., $w = w_{max}$). 
On the coefficients of the BGL expansions\,(\ref{eq:fbar_BGL}), truncated at orders $(N_g, N_f, N_{F_1}, 0)$, we impose the corresponding unitarity conditions\,(\ref{eq:UTfilter2}), adopting the values of the three reduced susceptibilities $\chi_{1^-, 1^+, 0^-} / f^2(1)$, obtained from the non-perturbative results\,(\ref{eq:bound1+})-(\ref{eq:f1}). The unitary conditions\,(\ref{eq:UTfilter2}) are fulfilled using the frequentist approach described in Appendix B of Ref.\,\cite{Simula:2023ujs}.

We generate a sample of ${\cal{O}}(10^4)$ events according to the multivariate Gaussian distributions of the experimental ratios $R_n(x)$, adopting for each experiment the corresponding covariance matrices $\mathbf{C}_{nm}$ and considering that the three datasets for the ratios $R_n(x)$ are uncorrelated (see also Refs.\,\cite{Banerjee:2024znd, FlavourLatticeAveragingGroupFLAG:2024oxs}).
For each event we apply the BGL expansions\,(\ref{eq:fbar_BGL}) for the hadronic FFs, truncated at orders $(N_g, N_f, N_{F_1}, 0)$, and we evaluate the helicity amplitudes given by Eqs.\,(\ref{eq:H+})-(\ref{eq:H0}). In this way, we obtain the SM predictions for the ratios $R_n(x)$ according to Eqs.\,(\ref{eq:ratio_w2})-(\ref{eq:Delta_Hii}) and\,(\ref{eq:eta})-(\ref{eq:epsilon}), using the experimental bins\,(\ref{eq:bins}). 
Then, we perform a $\chi^2$-minimization procedure based on a correlated $\chi^2$-variable\footnote{We have checked that the use of an uncorrelated $\chi^2$-variable in the minimization procedure leads to differences in the results well below $1 \sigma$.}. 
Since the covariance matrices $\mathbf{C}_{nm}$ are singular (see Appendix\,\ref{sec:bins}), we adopt the Moore-Penrose pseudoinverse matrices, commonly used in least-square procedures. 
The total number of all experimental data sets is $N_{data} = 40 + 40 + 38 = 118$ and the total number of free parameters is $N_{parms} = N_g + N_f + N_{F_1} + 1$. 
Since each of the matrices $\mathbf{C}_{nm}$ possesses 4 null eigenvalues, the total number of degrees of freedom is $N_{dof} = N_{data} - N_{parms} - 4 N_{exps}$, where $N_{exps} =3$ is the number of experimental data sets considered. 

\begin{figure}[htb!]
\begin{center}
\includegraphics[scale=0.35]{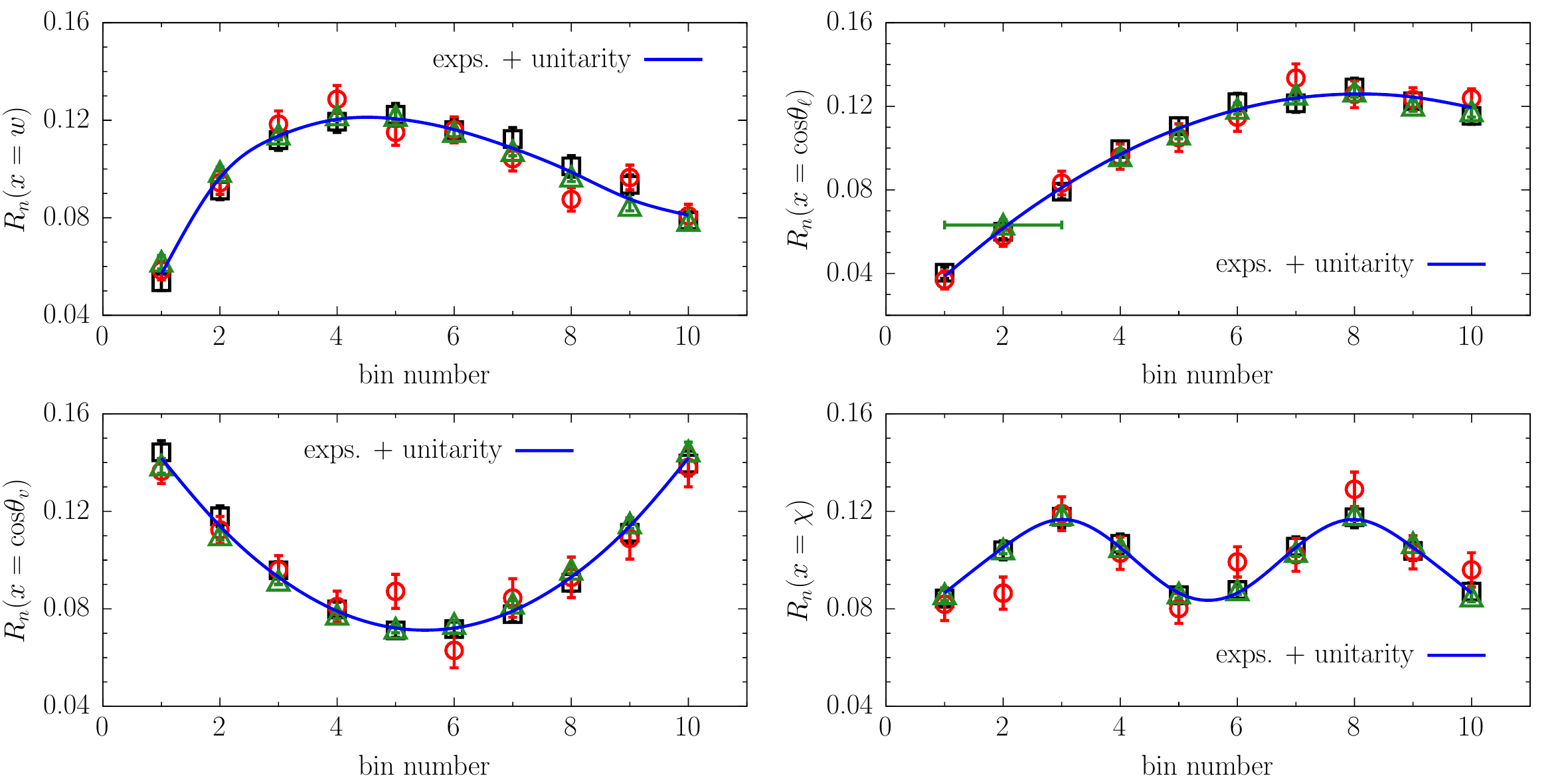}
\end{center}
\vspace{-0.75cm}
\caption{\it \small Comparison of the ratios $R_n(x)$, given by Eqs.\,(\ref{eq:ratios}), corresponding to the experiments Belle18\,\cite{Belle:2018ezy} (black squares), Belle23\,\cite{ Belle:2023bwv} (red circles) and BelleII23\,\cite{Belle-II:2023okj} (green triangles) with the results based on the unitary BGL fit\,(\ref{eq:fbar_BGL}) at orders $(2, 2, 2, 0)$. The value of the reduced $\chi^2$-variable is $\chi^2/N_{d.o.f.} \simeq 1.1$.}
\label{fig:ratios}
\end{figure}

We have employed various truncation orders $(N_g, N_f, N_{F_1}, 0)$ in Eqs.\,(\ref{eq:fbar_BGL}) and a quite good reproduction of the experimental ratios $R_n(x)$ is obtained already at orders $(2, 2, 2, 0)$. The quality of the fit is illustrated in Fig.\,\ref{fig:ratios}. Hereafter, we refer to our fitting procedure as ``exps. + unitarity" and we stress that such a procedure is different from joint fits of experimental and lattice data, often adopted in literature (see, e.g., Refs.\,\cite{Gambino:2019sif, Jaiswal:2020wer, Ray:2023xjn, Bordone:2024weh}). 
Indeed, the only theoretical ingredients needed in our SM analysis are the three reduced susceptibilities $\chi_{1^-, 1^+, 0^-} / f^2(1)$, which simply allow to implement the constraints of unitarity in defining the width of the band of values for the reduced FFs.

\begin{figure}[htb!]
\begin{center}
\includegraphics[scale=0.30]{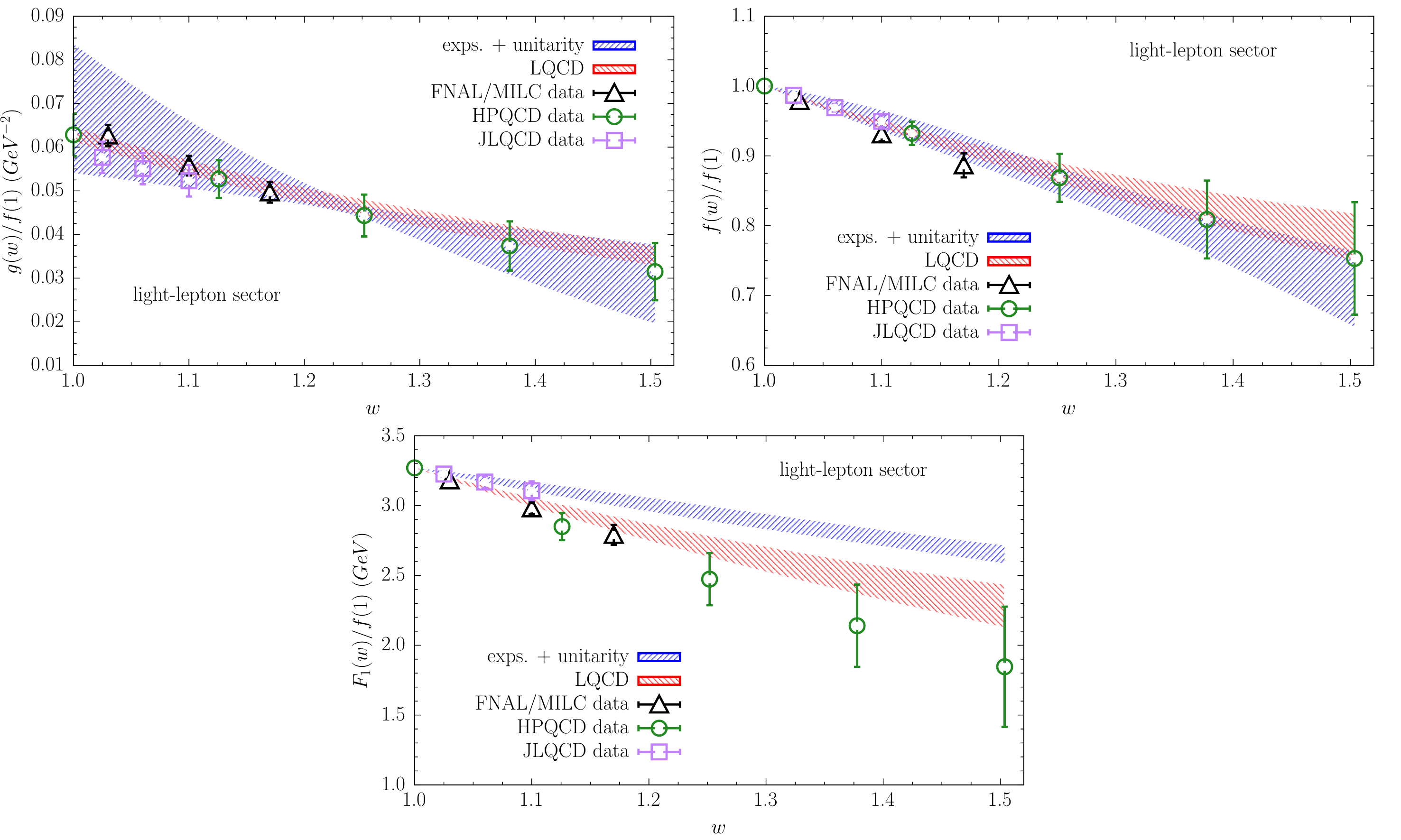}
\end{center}
\vspace{-0.5cm}
\caption{\it \small Comparison of the reduced FFs, extracted from a simultaneous analysis of the light-lepton experiments Belle18\,\cite{Belle:2018ezy}, Belle23\,\cite{ Belle:2023bwv}  and BelleII23\,\cite{Belle-II:2023okj}, with the individual LQCD FFs from FNAL/MILC\,\cite{FermilabLattice:2021cdg}, HPQCD\,\cite{Harrison:2023dzh} and JLQCD\,\cite{Aoki:2023qpa} Collaborations as well as with the results of the unitary DM method of Ref.\,\cite{Martinelli:2023fwm} based on all the LQCD FFs. The width of the blue and red bands correspond to one standard deviation.}
\label{fig:ffs_noPol}
\end{figure}

In Fig.\,\ref{fig:ffs_noPol} the $w$-dependence of the ``experimental" reduced FFs is shown and compared with the results of the individual LQCD simulations from FNAL/MILC\,\cite{FermilabLattice:2021cdg}, HPQCD\,\cite{Harrison:2023dzh} and JLQCD\,\cite{Aoki:2023qpa} Collaborations and with the results of the unitary DM method of Ref.\,\cite{Martinelli:2023fwm} based on all the LQCD FFs. In what follows, the latter case will be denoted simply as LQCD. For sake of precision, in order to construct the lattice results for the reduced FFs we have adopted the values $f(1) = 5.951 \pm 0.091$ GeV for FNAL/MILC, $f(1) = 5.885 \pm 0.094$ GeV for HPQCD and $f(1) = 5.776 \pm 0.090$ GeV for JLQCD, while for the case LQCD the value of $f(1)$ is given by Eq.\,(\ref{eq:f1}). While the value of $f(1)$ is directly given by HPQCD in Ref.\,\cite{Harrison:2023dzh}, those corresponding to FNAL/MILC and JLQCD have been calculated at zero recoil ($w = 1$) using the DM method of Ref.\,\cite{Martinelli:2023fwm} applied separately to the FNAL/MILC and JLQCD data points.

While both $\widetilde{f}$ and $\widetilde{g}$ appear to be consistent between ``exps. + unitarity" and LQCD, there are important differences in the slope of the FF $\widetilde{F}_1$ at a level similar to what occurs among the individual lattice results.
Since the latter differences have been discussed in Ref.\,\cite{Martinelli:2023fwm}, we now repeat the present analysis separately for each experimental data set. 

\begin{figure}[htb!]
\begin{center}
\includegraphics[scale=0.30]{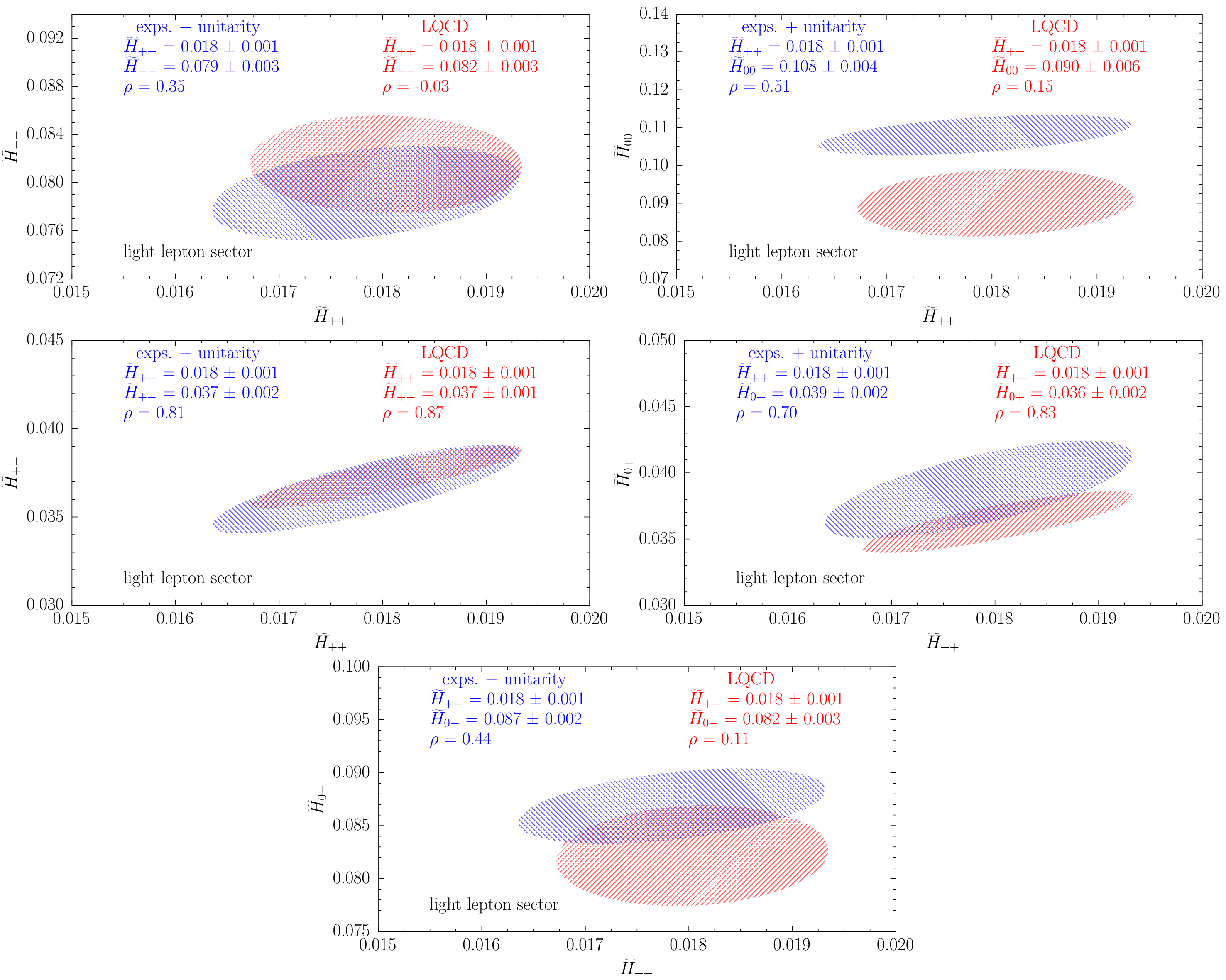} 
\end{center}
\vspace{-0.5cm}
\caption{\it \small Comparison of the contours at $68\%$ probability for the observables $\widetilde{H}_{ij} \equiv \overline{H}_{ij} / f^2(1)$, extracted from light-lepton experiments adopting the unitary BGL fit\,(\ref{eq:fbar_BGL}) at orders (2,2,2,0), with the corresponding LQCD predictions obtained by applying the unitary DM method of Ref.\,\cite{Martinelli:2023fwm} to the available LQCD results for the hadronic FFs from Refs.\,\cite{FermilabLattice:2021cdg, Harrison:2023dzh, Aoki:2023qpa}. In the insets the quantity $\rho$ represents the correlation coefficient.}
\label{fig:Hij}
\end{figure}

The results for the integrated observables $\widetilde{H}_{ij} \equiv \overline{H}_{ij} / f^2(1)$ (see Eq.\,(\ref{eq:Hij})) are collected in Table~\ref{tab:Hij} and visualized as contour plots in Fig.\,\ref{fig:Hij}, including the correlations among the various quantities.

\begin{table}[htb!]
\renewcommand{\arraystretch}{1.5}
\begin{center}
\begin{adjustbox}{max width=\textwidth}
\begin{tabular}{|c||c|c|c|c|c|c||}
\hline
& $\widetilde{H}_{++} (\%)$ & $\widetilde{H}_{--} (\%)$ & $\widetilde{H}_{00} (\%)$ & $\widetilde{H}_{+-} (\%)$ & $\widetilde{H}_{0+} (\%)$ & $\widetilde{H}_{0-} (\%)$ \\ \hline
\hline
Belle18                                                & 1.98 (25) & 8.43 (57) & 11.76 (87) & 3.94 (40) & 4.41 (62) & 9.37 (53) \\ \hline
Belle23                                                & 1.90 (22) & 8.11 (68) & 10.09 (86) & 3.75 (39) & 3.97 (57) & 8.20 (55) \\ \hline
BelleII23                                              & 2.02 (14) & 7.28 (34) & 10.39 (45) & 3.70 (22) & 3.98 (31) & 8.22 (31) \\ \hline \hline
Belle18 + Belle23 + BelleII23              & 1.78 (10) & 7.91 (26) & 10.80 (35) & 3.66 (17) & 3.87 (24) & 8.68 (23) \\ \hline \hline
LQCD                                                  & 1.80 ~(9) & 8.15 (27) & ~9.01 (59) & 3.73 (12) & 3.63 (16) & 8.22 (31) \\ \hline
\end{tabular}
\end{adjustbox}
\end{center}
\renewcommand{\arraystretch}{1.0}
\vspace{-0.25cm}
\caption{\it \small Results for the integrated observables $\widetilde{H}_{ij} \equiv \overline{H}_{ij} / f^2(1)$ (see Eq.\,(\ref{eq:Hij}), obtained using the reduced FFs extracted from the analysis of the ratios\,(\ref{eq:ratio_w2})-(\ref{eq:ratio_chi2}) corresponding separately to the Belle18\,\cite{Belle:2018ezy}, Belle23\,\cite{Belle:2023bwv} and BelleII23\,\cite{Belle-II:2023okj} experimental data sets. The  row denoted by Belle18+Belle23+BelleII23 corresponds to the results obtained using simultaneously all the three experimental data sets. The last row shows the SM predictions obtained using the hadronic FFs of the unitary DM approach of Ref.\,\cite{Martinelli:2023fwm} based on all available LQCD results from FNAL/MILC\,\cite{FermilabLattice:2021cdg}, HPQCD\,\cite{Harrison:2023dzh} and JLQCD\,\cite{Aoki:2023qpa} Collaborations.}
\label{tab:Hij}
\end{table}

It can be seen that:
\begin{itemize}

\item the integrated observables $\widetilde{H}_{ij}$, obtained using simultaneously all the three experimental data sets, deviates from the corresponding LQCD predictions less than one standard deviation, except the case of $\widetilde{H}_{00}$, which depends on the FF $\widetilde{F}_1$, where the difference reaches $\simeq 2.6 \sigma$;

\item all the results for $\widetilde{H}_{ij}$ corresponding to the experimental Belle23 data set agree with the LQCD results better than one standard deviation;

\end{itemize}

Similar results hold as well also for the asymmetries $A_{FB}$, $F_L$, $A_{1c}$, $A_{2c}$ and $A_{3c}$, given in Eqs.\,(\ref{eq:AFB})-(\ref{eq:A3c}), as described in Appendix\,\ref{sec:AFB}.

\begin{figure}[htb!]
\begin{center}
\includegraphics[scale=0.30]{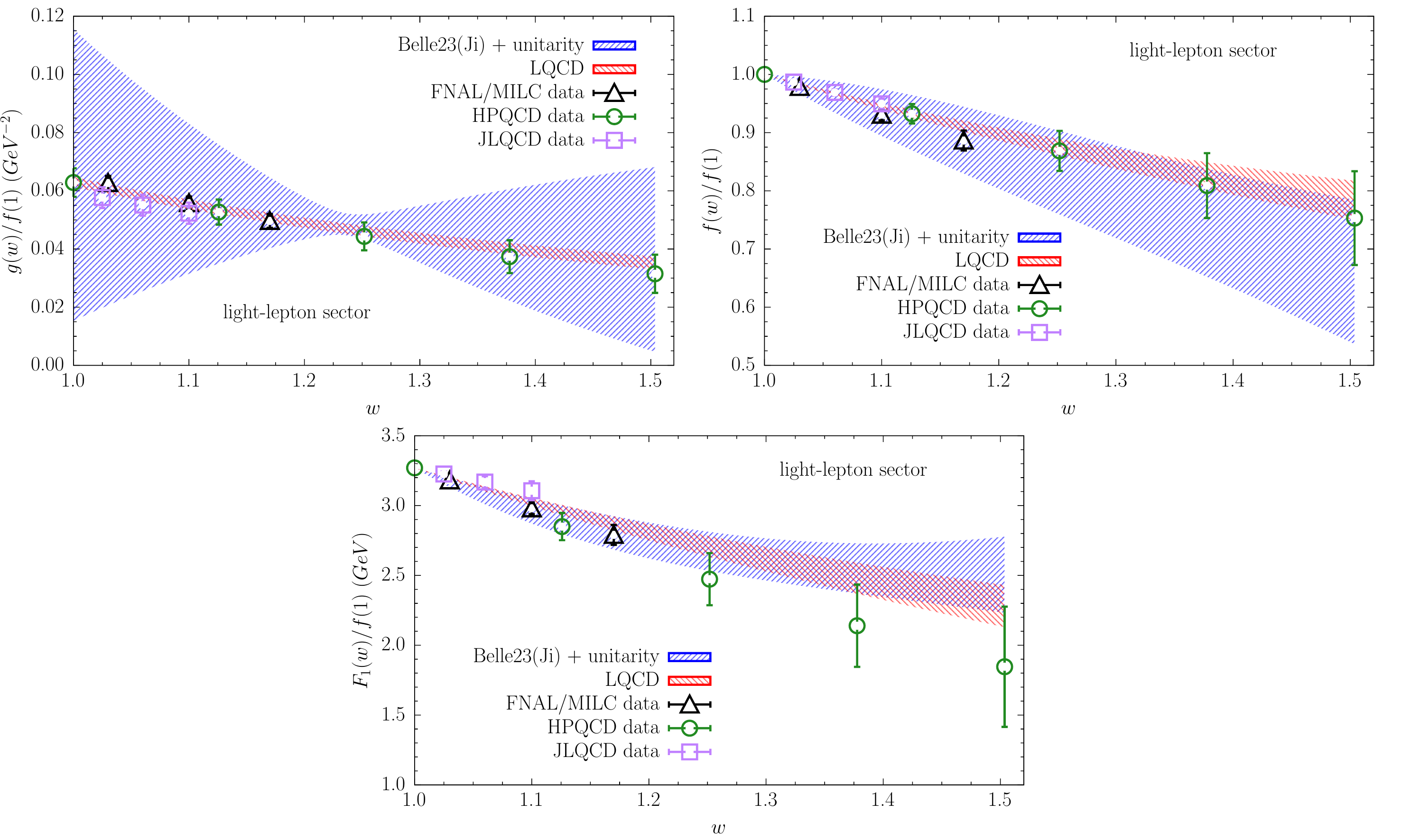}
\end{center}
\vspace{-0.5cm}
\caption{\it \small The same as in Fig.\,\ref{fig:ffs_noPol}, but for the reduced FFs extracted using the ratios\,(\ref{eq:ratio_w2})-(\ref{eq:ratio_chi2}) calculated using the complete set of angular coefficients of the four-fold differential decay rate, determined by the Belle Collaboration in Ref.\,\cite{Belle:2023xgj}, labelled as Belle23(Ji)\,\cite{Martinelli:2024vde}.}
\label{fig:ffs_Belle23_angular}
\end{figure}

Recently, the complete set of angular coefficients characterizing the four-fold differential decay rate has been determined in four $w$-bins by the Belle Collaboration in Ref.\,\cite{Belle:2023xgj}. 
Using the above results, we constructed in Ref.\,\cite{Martinelli:2024vde} a new data set for the ratios\,(\ref{eq:ratio_w2})-(\ref{eq:ratio_chi2}), labelled as Belle23(Ji). The two sets Belle23 and Belle23(Ji) are not independent, since they share the same four-fold differential data set, and they differ only in the way the single-differential decay rates are evaluated.
The analysis of the Belle23(Ji) data set carried out in Ref.\,\cite{Martinelli:2024vde} suggested that the $w$-dependence of the Belle23(Ji) data (and, therefore, also of the Belle23 set) may be compatible with the slope of the LQCD FFs.
To confirm and deepen the above suggestion, we have extracted the reduced FFs from either the Belle23 or the Belle23(Ji) data sets. The results obtained using the Belle23(Ji) data set are shown in Fig.\,\ref{fig:ffs_Belle23_angular} and compared with the individual LQCD simulations from FNAL/MILC\,\cite{FermilabLattice:2021cdg}, HPQCD\,\cite{Harrison:2023dzh} and JLQCD\,\cite{Aoki:2023qpa} Collaborations and with the results of the unitary DM method of Ref.\,\cite{Martinelli:2023fwm} based on all the LQCD FFs. Similar results hold as well for the Belle23 data set. 

Thus, an interesting conclusion of this Section is that the slope of the reduced form factor $\widetilde{F}_1(w) = F_1(w) / f(1)$ required by the Belle18+BelleII23 data set agrees well with the results of JLQCD Collaboration\,\cite{Aoki:2023qpa}, while the one required by either the Belle23 or the Belle23(Ji) data sets is consistent with the results from FNAL/MILC\,\cite{FermilabLattice:2021cdg} and HPQCD\,\cite{Harrison:2023dzh}. This finding confirms and improves the suggestion made in Ref.\,\cite{Martinelli:2024vde}. This analysis is very close, although with some differences, to the approaches followed in Refs\,\cite{Jaiswal:2017rve, Bernlochner:2017jka, Bigi:2017jbd, Bigi:2017njr, Gambino:2019sif, Jaiswal:2020wer, Li:2024weu}.

\section{Analysis of light-lepton experiments plus $\tau$-observables}
\label{sec:light+tau}

We now add in our SM analysis the $\tau$-observables described in Section\,\ref{sec:data} and given in Eqs.\,(\ref{eq:RDstar_exp})-(\ref{eq:Ptau_exp}). 
For completeness we have collected in Appendix\,\ref{sec:ratios_tau} the generalization of Eqs.\,(\ref{eq:ratio_w})-(\ref{eq:ratio_chi}) to the case of the massive $\tau$-lepton within the SM framework.

We have considered several cases, namely the addition of one specific $\tau$-observable at a time as well as the addition of several $\tau$-observables simultaneously. In the latter case the input data are treated as uncorrelated apart from the correlation between $(F_{L, \tau})_{<,>}$ measured by LHCb\,\cite{LHCb:2023ssl}(see text below Eq.\,\ref{eq:FLtau_high}), since no direct information is available from the experiments\footnote{We have checked the impact of possible correlations among $R(D^*)$ and $(F_{L, \tau})_{<,>}$, assuming a correlation coefficient equal to $\pm 0.5$. No significative effect has been found for the extracted reduced FFs.}. The only exception is represented by the experimental determinations of $F_{L, \tau}$ (see Eq.\,(\ref{eq:FLtau_exp})) and of $(F_{L, \tau})_{<,>}$ (see Eqs.\,(\ref{eq:FLtau_low})-(\ref{eq:FLtau_high})). Indeed, since the experimental value of $F_{L, \tau}$ include results from Ref.\,\cite{LHCb:2023ssl}, the quantities $F_{L, \tau}$ and $(F_{L, \tau})_{<,>}$ are not independent. Therefore, we never use them together and make use only of either $F_{L, \tau}$ or $(F_{L, \tau})_{<,>}$. Finally, as expected, the inclusion of the experimental value for the $\tau$-polarization $P_\tau(D^*)$ (see Eq.\,(\ref{eq:Ptau_exp})) has negligible effects on the extracted reduced FFs, because of the large uncertainty affecting its present determination\,(\ref{eq:Ptau_exp}).

The results of our fitting procedure are collected in Table\,\ref{tab:tau} and compared with the predictions based on the unitary DM method of Ref.\,\cite{Martinelli:2023fwm} applied to all LQCD data.
In Table\,\ref{tab:tau} we have included also the results for other two $\tau$-observables, namely the forward-backward asymmetry $A_{FB, \tau}$ and the transverse asymmetry $A_{1c, \tau}$ (their definitions in terms of the hadronic FFs can be easily inferred from, e.g., Ref.\,\cite{Martinelli:2024vde}).

\begin{figure}[htb!]
\begin{center}
\includegraphics[scale=0.35]{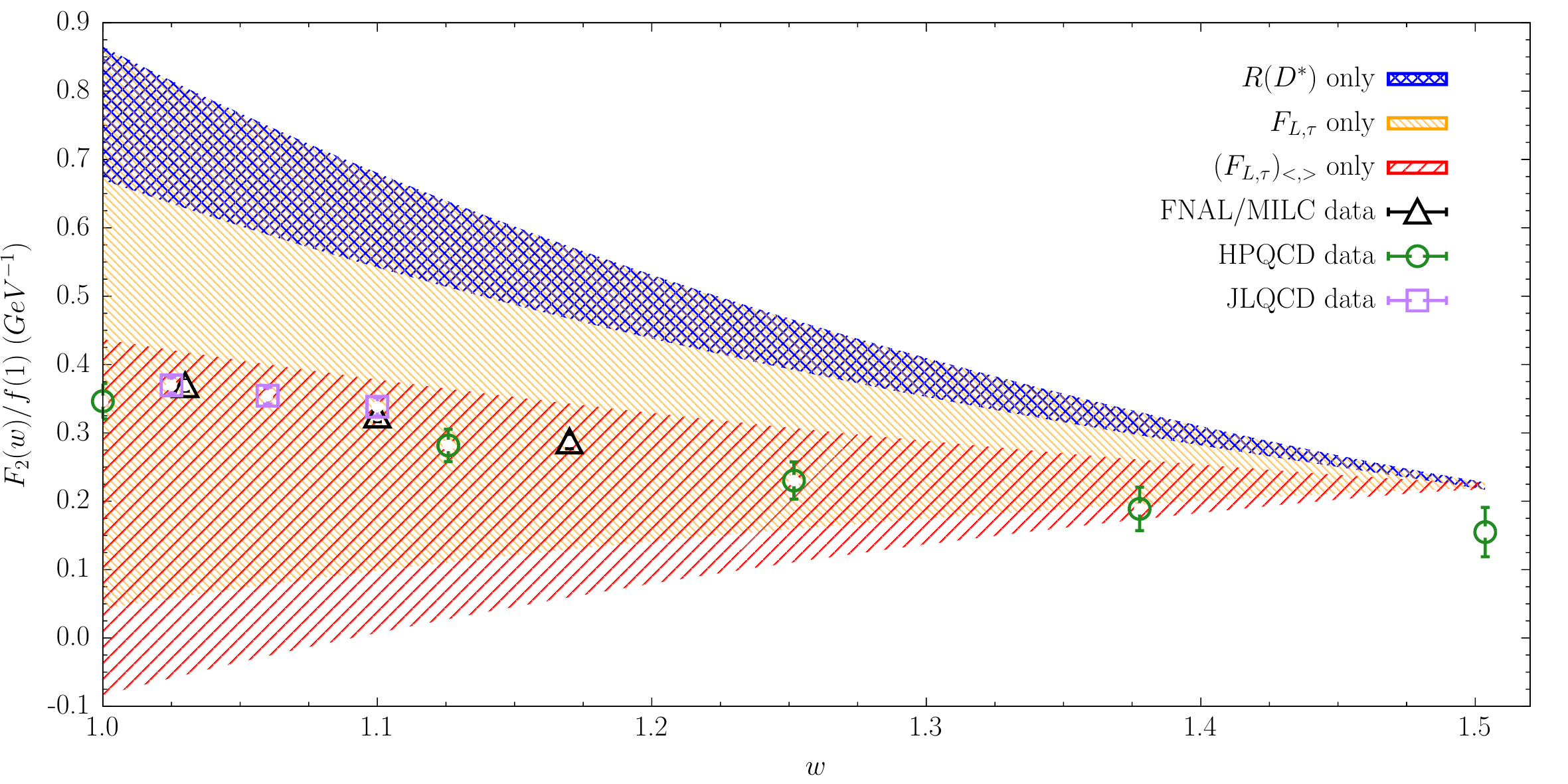}
\end{center}
\vspace{-0.5cm}
\caption{\it \small Comparison of the reduced FF $\widetilde{F}_2 = F_2 / f(1)$ extracted within the SM from the (light-lepton) experimental ratios $R_n(x)$ plus a single selected $\tau$-observable at a time with the individual LQCD results from FNAL/MILC\,\cite{FermilabLattice:2021cdg}, HPQCD\,\cite{Harrison:2023dzh} and JLQCD\,\cite{Aoki:2023qpa} Collaborations.}
\label{fig:F2_1}
\end{figure}

\begin{figure}[htb!]
\begin{center}
\includegraphics[scale=0.35]{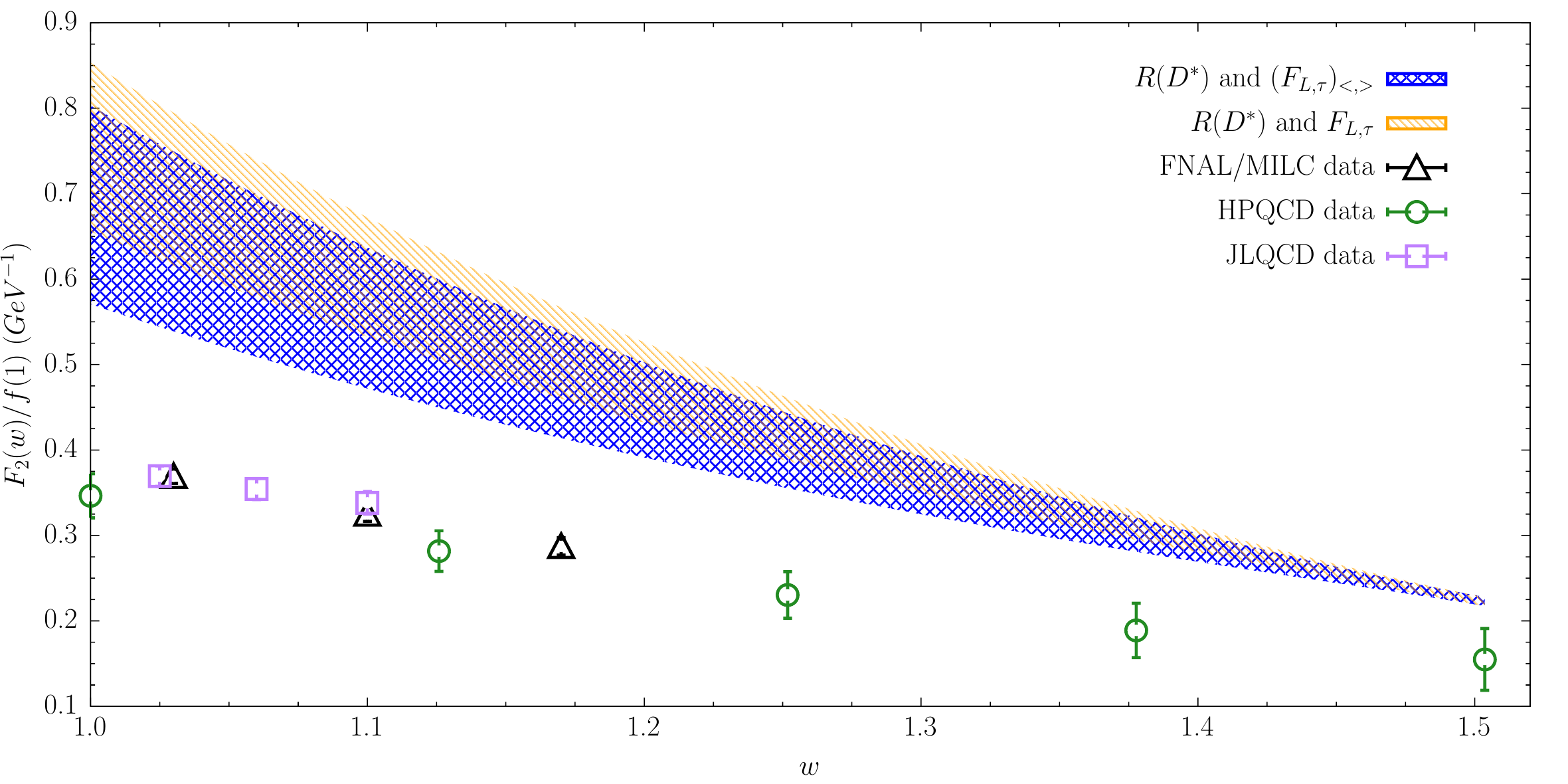}
\end{center}
\vspace{-0.5cm}
\caption{\it \small The same as in Fig.\,\ref{fig:F2_1}, but including in the fitting procedure besides the (light-lepton) experimental ratios $R_n(x)$ also the experimental values of $R(D^*)$ and of either $F_{L, \tau}$ or $(F_{L, \tau})_{<,>}$.}
\label{fig:F2_2}
\end{figure}

The extracted reduced FFs $\widetilde{g}$, $\widetilde{f}$ and $\widetilde{F}_1$ are dominated by the light-lepton data and basically coincides with those determined in Section\,\ref{sec:light} (see Fig.\,\ref{fig:ffs_noPol}). The $w$-dependence of the extracted reduced FF $\widetilde{F}_2$ is shown in Figs.\,\ref{fig:F2_1} and\,\ref{fig:F2_2} and compared with the results of the individual LQCD simulations from FNAL/MILC\,\cite{FermilabLattice:2021cdg}, HPQCD\,\cite{Harrison:2023dzh} and JLQCD\,\cite{Aoki:2023qpa} Collaborations.

\begin{table}[htb!]
\renewcommand{\arraystretch}{1.5}
\begin{center}
\begin{adjustbox}{max width=\textwidth}
\begin{tabular}{|c||c||c|c|c||c||c||c||}
\hline
& $R(D^*)$ & $F_{L, \tau}$ & $(F_{L, \tau})_>$ &  $(F_{L, \tau})_<$ & $P_\tau(D^*)$ & $A_{FB, \tau}$ & $A_{1c, \tau}$ \\ \hline
\hline
experiment                                          & 0.286 (12) & 0.48 (9) & 0.34 (8) & 0.52 (8) & -0.36 ~(54) & -- & -- \\ \hline \hline
only $a_0^{F_2}$                                & 0.248 ~(1) & 0.446 ~(4) &  0.394 ~(5) & 0.536 ~(4) &-0.524 ~(4) & ~0.073 (15) & -0.110 ~(4) \\ \hline \hline
with $R(D^*)$                                      & 0.284 (10) & 0.515 (18) &  0.451 (16) & 0.616 (19) & -0.333 (48) & -0.012 (22) & -0.096 ~(5) \\ \hline
with $F_{L, \tau}$                                & 0.265 (26) & 0.474 (52) &  0.420 (40) & 0.564 (66) & -0.44 ~(14) & ~0.056 (88) & -0.104 (11) \\ \hline
with $(F_{L, \tau})_{<,>}$                    & 0.243 (14) & 0.432 (30) &  0.386 (22) & 0.513 (41) & -0.561 (83) &~0.114 (61) &-0.113 ~(7) \\ \hline \hline
with $R(D^*)$,  $F_{L, \tau}$              & 0.283 (10) & 0.513 (18) &  0.449 (16) & 0.614 (19) & -0.339 (48) & -0.010 (22) & -0.097 ~(5) \\ \hline
with $R(D^*)$,  $(F_{L, \tau})_{<,>}$  & 0.277 (11) & 0.500 (20) &  0.438 (18) & 0.600 (23) & -0.372 (55) &~0.003 (26) & -0.099 ~(5) \\ \hline \hline
LQCD                                                  & 0.258 ~(5) & 0.426 ~(8) &  0.384 ~(5) & 0.499 (12) & -0.521 ~(6) & ~0.078 ~(8) &-0.115 ~(2) \\ \hline
\end{tabular}
\end{adjustbox}
\end{center}
\renewcommand{\arraystretch}{1.0}
\vspace{-0.25cm}
\caption{\it \small Experimental values of $R(D^*)$\,\cite{Banerjee:2024znd}, $F_{L, \tau}$\,\cite{Belle:2019ewo, LHCb:2023ssl}, $(F_{L, \tau})_{<,>}$\,\cite{LHCb:2023ssl} and $P_\tau(D^*)$\,\cite{Belle:2016dyj} compared with the results obtained within the SM using the unitary BGL fit\,(\ref{eq:fbar_BGL}) of the (light-lepton) experimental ratios $R_n(x)$ plus selected $\tau$-observables. The last two columns contain the corresponding predictions for the forward-backward asymmetry $A_{FB, \tau}$ and the transverse asymmetry $A_{1c, \tau}$, respectively. The third row corresponds to the case in which no $\tau$-observables are included in the fitting procedure; in this case the value of the BGL coefficient $a_0^{F_2}$ is simply fixed by the KC\,(\ref{eq:KC2}). For all fits the value of the reduced $\chi^2$-variable is $\chi^2 / N_{d.o.f.} \simeq 1.1 - 1.2$.The last row shows the LQCD predictions of the unitary DM method of Ref.\,\cite{Martinelli:2023fwm} applied to all available LQCD simulations of the hadronic FFs from FNAL/MILC\,\cite{FermilabLattice:2021cdg}, HPQCD\,\cite{Harrison:2023dzh} and JLQCD\,\cite{Aoki:2023qpa} Collaborations. The LQCD results correspond to the ones labelled ``Combined" in Table 4 and 5 of Ref.\,\cite{Martinelli:2023fwm}.}
\label{tab:tau}
\end{table}

The following comments are in order.

\begin{itemize}

\item The inclusion of the experimental values of either $F_{L, \tau}$ or $(F_{L, \tau})_{<,>}$ (without $R(D^*)$) produces a reduced FF $\widetilde{F}_2$, which is overall consistent with the LQCD results. Note that the uncertainty of the extracted FF $\widetilde{F}_2$, clearly visible in Fig.\,\ref{fig:F2_1}, is presently dominated by the large uncertainty in the experimental values of $F_{L, \tau}$ or $(F_{L, \tau})_{<,>}$ (see Eqs.\,(\ref{eq:FLtau_exp}-(\ref{eq:FLtau_high})). On the contrary, the inclusion of the experimental value of $R(D^*)$ (with or without the addition of $F_{L, \tau}$ or $(F_{L, \tau})_{<,>}$) demands a FF $\widetilde{F}_2$ significantly larger than the LQCD calculations by a factor $\sim 1.5 - 2$, i.e.~well beyond the differences among the individual LQCD results.

\item The large value of $\widetilde{F}_2$, required by the  inclusion of the experimental value of $R(D^*)$ (either with or without $(F_{L, \tau})_{<,>}$) produces values of various $\tau$-observables, which may differ significantly with respect to the corresponding LQCD predictions, namely (see, e.g., the last two rows of Table\,\ref{tab:tau}): $\simeq 1.5\sigma$ for $R(D^*)$, $\simeq 3.3 \sigma$ for $F_{L, \tau}$, $\simeq 2.8 \sigma$ for $(F_{L, \tau})_>$, $\simeq 3.9 \sigma$ for $(F_{L, \tau})_<$, $\simeq 2.6 \sigma$ for $P_\tau(D^*)$, $\simeq 2.7 \sigma$ for $A_{FB, \tau}$ and $\simeq 3.0 \sigma$ for $A_{1c, \tau}$.

\item The precision of the values of the $\tau$-observables other than $R(D^*)$, calculated using the extracted reduced FFs, is significantly better than the one of the corresponding experimental values.

\item The inclusion of the experimental value of $R(D^*)$ in our fitting procedure dominates over the addition of the experimental values of the other $\tau$-observables.

\end{itemize}

The above findings indicates a difficulty in reproducing simultaneously the experimental values of $R(D^*)$ and of the other $\tau$-observables within the SM. In particular, we have shown the presence of an important correlation among $R(D^*)$ and $(F_{L,\tau})_{<,>}$ and of a relevant anti-correlation between $R(D^*)$ and $A_{FB, \tau}$, a quantity not yet measured.
In Ref.\,\cite{Fedele:2023ewe} it was noticed that also $F_{L, \mu}$ is correlated with $R(D^*)$ (see also Ref.\,\cite{Gambino24}). It is worth highlighting that, while the latter effect is driven only by the FF $\widetilde{F}_1(w)$, the former one is determined by both $\widetilde{F}_1(w)$ and $\widetilde{F}_2(w)$.

\begin{figure}[htb!]
\begin{center}
\includegraphics[scale=0.30]{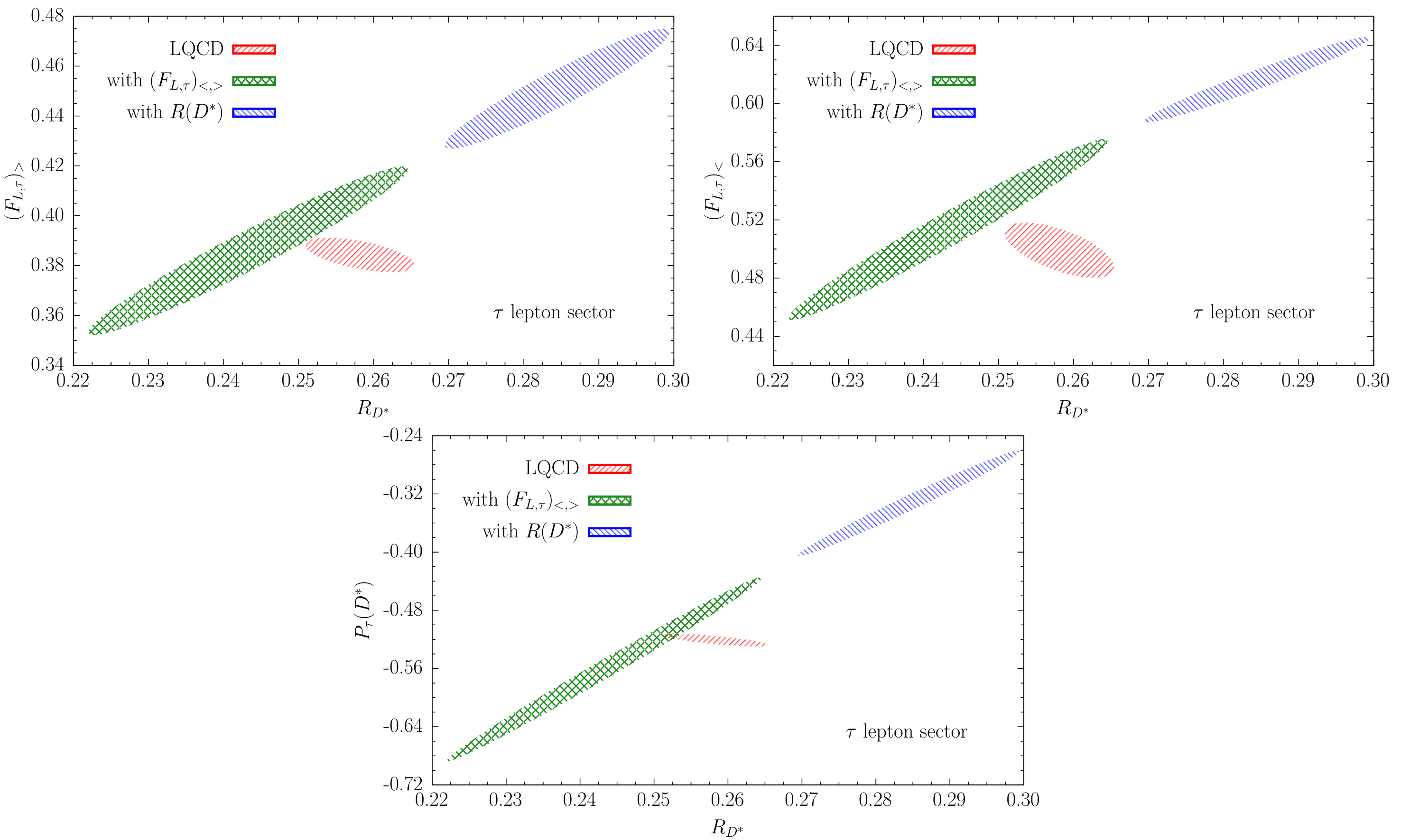}
\end{center}
\vspace{-0.5cm}
\caption{\it \small Comparison of the contours at $68\%$ probability for the observables $R(D^*)$, $(F_{L, \tau})_{<,>}$ and $P_\tau(D^*)$, extracted within the SM from the light-lepton experiments plus the experimental values of either $R(D^*)$ or $(F_{L, \tau})_{<,>}$, with the corresponding LQCD predictions, obtained by the unitary DM method of Ref.\,\cite{Martinelli:2023fwm} applied to all available LQCD FFs from Refs.\,\cite{FermilabLattice:2021cdg, Harrison:2023dzh, Aoki:2023qpa}. }
\label{fig:contour_tau}
\end{figure}

The above issue concerning the consistency among various $\tau$-observables is strengthened in Fig.\,\ref{fig:contour_tau}, where we show the contour plots for the observables $R(D^*)$, $(F_{L, \tau})_{<,>}$ and $P_\tau(D^*)$, extracted within the SM from the light-lepton experiments plus the experimental values of either $R(D^*)$ or $(F_{L, \tau})_{<,>}$, and those corresponding to the LQCD predictions, obtained by the unitary DM method of Ref.\,\cite{Martinelli:2023fwm} applied to all LQCD data.

Thus, any improvement in the current precision of the experimental determinations of the $\tau$-observables as well as a first-time determination of the forward-backward $A_{FB, \tau}$ and transverse $A_{1c, \tau}$ asymmetries, will be crucial to learn more on the suggested inconsistency found  within the SM between the experimental values of $R(D^*)$ and of the other $\tau$-observables.

 In order to use our results for further phenomenological analyses we have collected in Appendix\,\ref{sec:corr} the mean values with uncertainties and the correlation matrix for various light- and $\tau$-lepton observables. Our results are shown in Tables\,\ref{tab:correxp} and\,\ref{tab:corrth} respectively for the last two cases of Table\,\ref{tab:tau}.
 
Before closing this Section, we consider the so-called HQET-inspired FFs $h_V(w)$, $h_{A_1}(w)$, $h_{A_2}(w)$ and $h_{A_3}(w)$, related to the BGL-like FFs $g(w)$, $f(w)$, $F_1(w)$ and $F_2(w)$ by
\bea
      \label{eq:hV}
      h_V(w) & = & m_B \sqrt{r} g(w) ~ , ~ \nonumber \\[2mm]
      \label{eq:hA1}
      h_{A_1}(w) & = & \frac{1}{m_B \sqrt{r} (1 + w)} f(w) ~ , ~ \\[2mm]
      \label{eq:hA2}
      h_{A_2}(w) & = & \frac{1}{m_B^2 \sqrt{r} (1 + w) (1 + r^2 - 2 r w)} \Bigl\{ \frac{(w -r) F_1(w) - m_B (1 + r^2 - 2 r w) f(w)}{w-1} \nonumber \\[2mm]
                         & - & m_B^2 r (1 + w) F_2(w) \Bigl\} ~ , ~ \nonumber \\[2mm]
      \label{eq:hA3}
      h_{A_3}(w) & = & \frac{1}{m_B \sqrt{r} (1 + r^2 - 2 r w)} \Bigl\{ \frac{1 - w r}{w^2 -1} \left[ (w -r) f(w) - \frac{F_1(w)}{m_B}\right] \nonumber \\[2mm]
                         & + & m_B r^2 F_2(w) -r f(w) \Bigl\} ~ . ~ \nonumber
\eea

In the limit of infinite heavy-quark masses one has simply $h_V(w) = h_{A_1}(w) = h_{A_3}(w) = \xi_{IW}(w)$, while $h_{A_2}(w) = 0$, where $\xi_{IW}(w)$ is the universal, leading Isgur-Wise function\,\cite{Isgur:1989vq, Isgur:1990yhj}.
We remind also that the HQET FFs should not satisfy any KCs at variance with the FFs having definite spin-parity.

\begin{figure}[htb!]
\begin{center}
\includegraphics[scale=0.30]{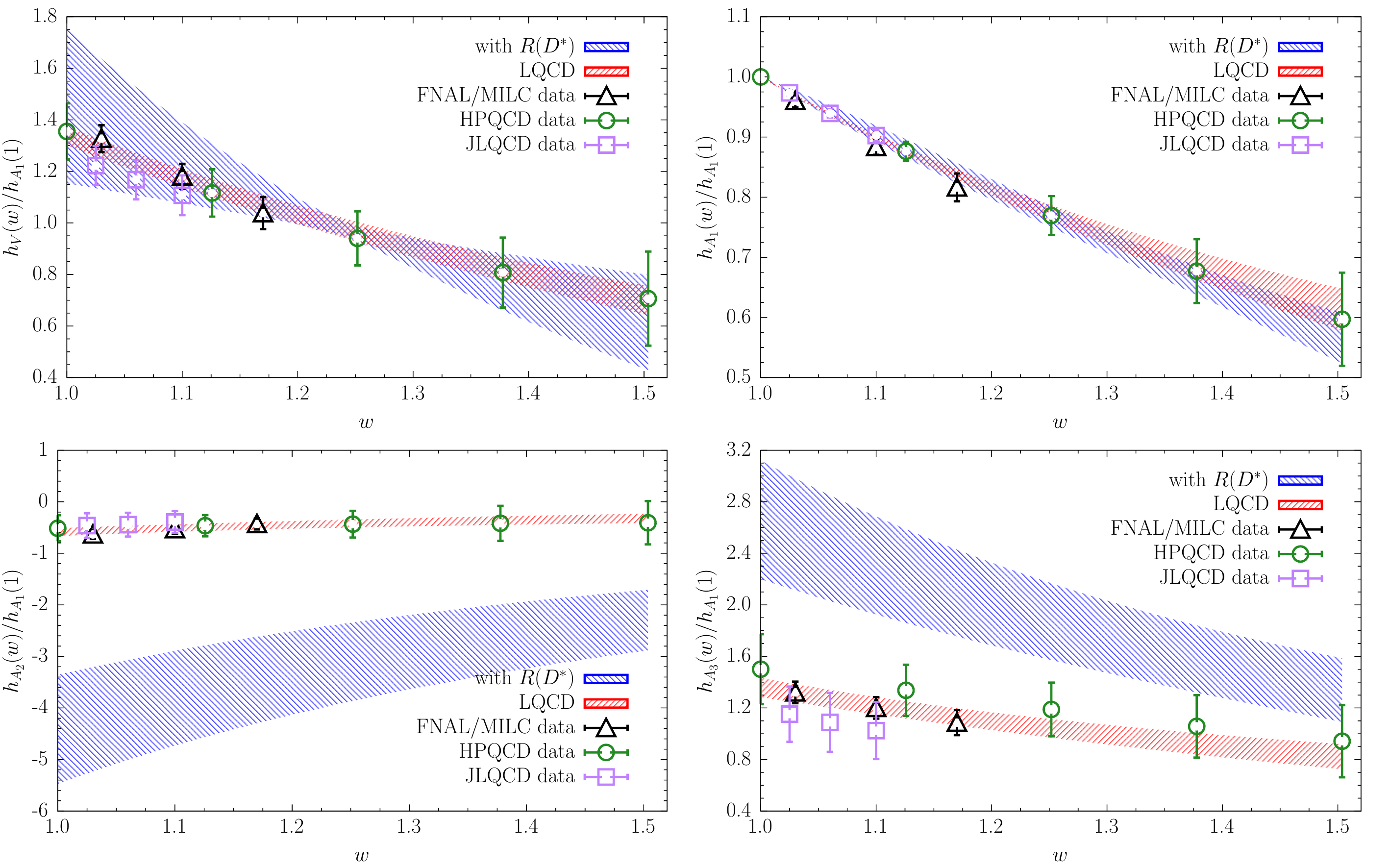}
\end{center}
\vspace{-0.5cm}
\caption{\it \small Comparison of the HQET-inspired FFs $h_V(w)$, $h_{A_1}(w)$, $h_{A_2}(w)$ and $h_{A_3}(w)$, divided by $h_{A_1}(1) = f(1) / (2 m_B \sqrt{r})$, extracted within the SM from the (light-lepton) experimental ratios $R_n(x)$ plus the experimental value of $R(D^*)$, with the individual LQCD results from FNAL/MILC\,\cite{FermilabLattice:2021cdg}, HPQCD\,\cite{Harrison:2023dzh} and JLQCD\,\cite{Aoki:2023qpa} Collaborations, as well as with the LQCD predictions, obtained by the unitary DM method of Ref.\,\cite{Martinelli:2023fwm} applied to all available LQCD data.}
\label{fig:HQET}
\end{figure}

In Fig.\,\ref{fig:HQET} we show the HQET FFs\,(\ref{eq:hA1}), divided by $h_{A_1}(1) = f(1) / (2 m_B \sqrt{r})$, corresponding to the reduced FFs extracted from the (light-lepton) experimental ratios $R_n(x)$ plus the experimental value of $R(D^*)$. The comparison with the individual LQCD results from FNAL/MILC\,\cite{FermilabLattice:2021cdg}, HPQCD\,\cite{Harrison:2023dzh} and JLQCD\,\cite{Aoki:2023qpa} Collaborations, as well as with the LQCD predictions, obtained by the unitary DM method of Ref.\,\cite{Martinelli:2023fwm} applied to all available LQCD data, indicates that the experimental value of $R(D^*)$ would require huge deviations from the heavy-quark limit for the FFs $h_{A_2}(w)$ and $h_{A_3}(w)$.

For the sake of completeness, in Appendix\,\ref{sec:Ri} we present our results for the ratios of the HQET FFs\,(\ref{eq:hA1}) with respect to $h_{A_1}(w)$.

\section{Results for $|V_{cb} f(1)|$}
\label{sec:Vcbf1}

In this Section we briefly present our results for product of the CKM matrix element $|V_{cb}|$ and the FF $f(1)$, $|V_{cb} f(1)|$.
From Eq.\,(\ref{eq:Gamma_ell}) the total decay rate for massless leptons can be rewritten as
\be
     \label{eq:Gamma_FFs}
     \Gamma =  \frac{4}{3} \frac{\eta_{EW}^2 m_B m_{D^*}^2}{(4 \pi)^3} G_F^2 |V_{cb} f(1)|^2 \left[ \widetilde{H}_{++} + \widetilde{H}_{--} + \widetilde{H}_{00} \right] ~ , ~
\ee
where the quantities $\widetilde{H}_{++, --, 00}  \equiv \overline{H}_{++, --, 00}  / f^2(1)$ can be calculated using a set of reduced FFs $\widetilde{f}$, $\widetilde{g}$ and $\widetilde{F}_1$.
Thus, the value of $|V_{cb} f(1)|$ can be obtained using the experimental value of total decay rate, $\Gamma_{exp}$, in the l.h.s.\,of Eq.\,(\ref{eq:Gamma_FFs}) and a given set of hadronic FFs in its r.h.s.

As for $\Gamma_{exp}$ we take from the latest PDG review\,\cite{ParticleDataGroup:2024cfk} the branching ratios $\mbox{Br}(B^0 \to D^* \ell \nu_\ell) = 4.90 (12) \%$ and $\mbox{Br}(B^+ \to D^* \ell \nu_\ell) = 5.53 (22) \%$, and the mean lifetimes $\tau_{B^0} = 1.517 (4) \cdot 10^{-12}$  s and $\tau_{B^+} = 1.638 (4) \cdot 10^{-12}$ s. After averaging the branching ratio divided by the mean lifetime in the two isospin channels, we get $\Gamma_{exp} = 21.74 (51) \cdot 10^{-15}$ GeV.

In the two Tables~\ref{tab:Vcb} and \ref{tab:Vcb_LQCD} we present our results for $|V_{cb} f(1)|$ adopting two different sets of hadronic FFs.

In Table\,\ref{tab:Vcb} we consider the reduced FFs extracted from the analysis of the (light-lepton) experimental ratios $R_n(x)$ (see Fig.\,\ref{fig:ffs_noPol}). 
Adopting for $f(1)$ the value $f(1) = 5.845 (50)$ GeV (see Eq.\,(\ref{eq:f1})), obtained in Ref.\,\cite{Martinelli:2023fwm} by the DM method using as inputs all available LQCD results, we get the estimates of $|V_{cb}|$ shown in the last column of Table\,\ref{tab:Vcb}. Our findings are nicely consistent with the result $|V_{cb}| = 39.92 (64) \cdot 10^{-3}$, obtained in Ref.\,\cite{Martinelli:2023fwm}, and with the value $|V_{cb}| = 39.46 (53) \cdot 10^{-3}$, quoted by the FLAG review\,\cite{FlavourLatticeAveragingGroupFLAG:2024oxs}. They however deviate by $\approx 2.6 \sigma$ from the latest determination $|V_{cb}| = 42.00 (47) \cdot 10^{-3}$\,\cite{Fael:2024rys} from inclusive semileptonic decays and  by $\approx 2.7 \sigma$ from the updated prediction $|V_{cb}| = 42.22 (51) \cdot 10^{-3}$ obtained within the SM by UTfit\,\cite{UTfit:2022hsi}.

\begin{table}[htb!]
\renewcommand{\arraystretch}{1.5}
\begin{center}
\begin{adjustbox}{max width=\textwidth}
\begin{tabular}{|c||c|c||}
\hline
& ~ $|V_{cb} f(1))| \cdot 10^3 ~ \mbox{GeV}^{-1}$ ~ & $ ~ |V_{cb}| \cdot 10^3 ~$ \\ \hline
\hline
Belle18                                                & 222.9 (8.5) & 38.1 (1.5) \\ \hline
Belle23                                                & 234.2 (9.3) & 40.1 (1.6) \\ \hline
BelleII23                                              & 236.5 (5.6) & 40.5 (1.0) \\ \hline \hline
Belle18 + Belle23 + BelleII23              & 231.8 (4.6) & 39.7 (0.8) \\ \hline \hline
\end{tabular}
\end{adjustbox}
\end{center}
\renewcommand{\arraystretch}{1.0}
\vspace{-0.25cm}
\caption{\it \small Results for the quantity $|V_{cb} f(1)|$ obtained (see text) adopting the experimental total decay rate $\Gamma_{exp} = 21.74 (51) \cdot 10^{-15}$ GeV from PDG\,\cite{ParticleDataGroup:2024cfk} and using the reduced FFs extracted from the analysis of the ratios $R_n(x)$ corresponding separately to the Belle18\,\cite{Belle:2018ezy}, Belle23\,\cite{Belle:2023bwv} and BelleII23\,\cite{Belle-II:2023okj} experimental data sets. The row denoted by Belle18+Belle23+BelleII23 corresponds to the results obtained using simultaneously all the three experimental data sets. The last column shows the value of $|V_{cb}|$ obtained using for $f(1)$ the value\,(\ref{eq:f1}).}
\label{tab:Vcb}
\end{table}

In Table\,\ref{tab:Vcb_LQCD} instead we adopt the hadronic FFs obtained in Ref.\,\cite{Martinelli:2023fwm} by the DM method using as inputs the LQCD results from FNAL/MILC\,\cite{FermilabLattice:2021cdg}, HPQCD\,\cite{Harrison:2023dzh} and JLQCD\,\cite{Aoki:2023qpa} Collaborations. 
Using for $f(1)$ the value corresponding to each  LQCD input (and shown in the third column of Table\,\ref{tab:Vcb_LQCD}), the resulting values of $|V_{cb}|$ are shown in last column of Table\,\ref{tab:Vcb_LQCD}. 
\begin{table}[htb!]
\renewcommand{\arraystretch}{1.5}
\begin{center}
\begin{adjustbox}{max width=\textwidth}
\begin{tabular}{|c||c|c||c||}
\hline
& ~ $|V_{cb} f(1))| \cdot 10^3 ~ \mbox{GeV}^{-1}$ ~ & ~ $f(1) ~ \mbox{GeV}^{-1}$ ~ & $ ~ |V_{cb}| \cdot 10^3 ~$ \\ \hline
\hline
FNAL/MILC  & 253.2 ~(9.2) & 5.951 (91) & 42.6 (1.7) \\ \hline
HPQCD        & 253.4 (11.7) & 5.885 (94) & 43.1 (2.1) \\ \hline
JLQCD         & 231.3 ~(9.5) & 5.776 (90) & 40.0 (1.8) \\ \hline \hline
LQCD           & 243.2 ~(5.9) & 5.845 (50) & 41.6 (1.1) \\ \hline \hline
\end{tabular}
\end{adjustbox}
\end{center}
\renewcommand{\arraystretch}{1.0}
\vspace{-0.25cm}
\caption{\it \small Results for the quantity $|V_{cb} f(1)|$ obtained (see text) adopting the experimental total decay rate $\Gamma_{exp} = 21.74 (51) \cdot 10^{-15}$ GeV from PDG\,\cite{ParticleDataGroup:2024cfk} and using the hadronic FFs obtained in Ref.\,\cite{Martinelli:2023fwm} by the DM method using as inputs the LQCD results from FNAL/MILC\,\cite{FermilabLattice:2021cdg}, HPQCD\,\cite{Harrison:2023dzh} and JLQCD\,\cite{Aoki:2023qpa} Collaborations. The row denoted by LQCD corresponds to the FFs obtained in Ref.\,\cite{Martinelli:2023fwm} by the DM method using all available LQCD FFs. The third column shows the value of $f(1)$ corresponding to the various sets of FFs, while  the corresponding values of $|V_{cb}|$ are presented in the last column.}
\label{tab:Vcb_LQCD}
\end{table}
Within the uncertainties the results for $|V_{cb}|$ in Tables\,\ref{tab:Vcb} and\,\ref{tab:Vcb_LQCD} are consistent.
Note that the use of the LQCD FFs to estimate the combination $\left( \widetilde{H}_{++} + \widetilde{H}_{--} + \widetilde{H}_{00} \right)$, appearing in the r.h.s.~of Eq.\,(\ref{eq:Gamma_FFs}), leads to larger central values of $|V_{cb}|$, which are consistent within one standard deviation with the latest inclusive determination $|V_{cb}| = 42.00 (47) \cdot 10^{-3}$\,\cite{Fael:2024rys} as well as with the updated prediction $|V_{cb}| = 42.22 (51) \cdot 10^{-3}$ obtained within the SM by the UTfit Collaboration in Ref.\,\cite{UTfit:2022hsi}. We stress that, in order to obtain the UTfit prediction for $\vert V_{cb} \vert$, the corresponding value of $\vert V_{cb} \vert$ labelled as measurement in Table I of Ref.\,\cite{UTfit:2022hsi} is not used at all in the global fit, so that the value $\vert V_{cb} \vert = 42.22(51) \cdot 10^{-3}$ is derived as a prediction, completely independent of any experimental and theoretical information on $B$-meson semileptonic decays, as described in details in Ref.\,\cite{UTfit:2022hsi}.

We finally comment on the difference between the result $|V_{cb}| = 39.92 (64) \cdot 10^{-3}$ of Ref.\,\cite{Martinelli:2023fwm} and the value $|V_{cb}| = 41.6 (1.1) \cdot 10^{-3}$ of Table\,\ref{tab:Vcb_LQCD}. Both determinations are based on the use of the same LQCD FFs.  They however rely on different data: on the one hand the differential distributions data and on the other hand the total decay rate data. 
We suggest that the reason may be traced back to the difference in the momentum dependence of the FF $F_1(w)$ visible in Fig.\,\ref{fig:ffs_noPol} (see also Refs.\,\cite{FermilabLattice:2021cdg, Martinelli:2021myh, Bordone:2024weh}).

\section{Conclusions}
\label{sec:conclusions}

We have presented an analysis of semileptonic  $B \to D^* \ell \nu_\ell$ decays based on the SM. We have extended our previous work\,\cite{Martinelli:2024vde}, limited to light leptons in the final state, by including also semitauonic decays.

Besides the experimental data on light leptons from Refs.\,\cite{Belle:2018ezy, Belle:2023bwv, Belle-II:2023okj} we have considered the inclusion of the experimental values of the ratio of branching fractions $R(D^*)$, the longitudinal $D^*$-polarization fraction $F_{L, \tau}$ and the $\tau$-polarization $P_\tau(D^*)$. For all lepton channels we make use of quantities that can be analyzed without the knowledge of $|V_{cb}|$. The hadronic FFs relevant in the SM are extracted directly from the experimental data using a fitting procedure based on the standard BGL $z$-expansions of the FFs with built-in constraints from unitarity. The only theoretical inputs are the susceptibilities required to impose the constraints of unitarity on the extracted FFs. 

The main results of our work are listed in the Introduction and will be not repeated here.

The results of our analysis indicates a difficulty in reproducing simultaneously the experimental values of $R(D^*)$ and of other $\tau$-observables within the SM. In particular, we found an important correlation among $R(D^*)$ and $(F_{L,\tau})_{<,>}$ and a relevant anti-correlation between $R(D^*)$ and $A_{FB, \tau}$, a quantity not yet measured.

We stress the importance of improving the current precision of the measurements of the $\tau$-observables as well as of performing a first-time determination of the forward-backward $A_{FB, \tau}$ and transverse $A_{1c, \tau}$ asymmetries. This will be crucial to learn more on the suggested inconsistency found within the SM between the experimental values of $R(D^*)$ and of other $\tau$-observables.

\section*{Acknowledgements}
S.S.~is supported by the Italian Ministry of Research (MIUR) under grant PRIN 2022N4W8WR.
The work of L.V.~is supported by the French Agence Nationale de la Recherche (ANR) under contracts ANR-19-CE31-0016 (`GammaRare') and ANR-23-CE31-0018 (`InvISYble').

\appendix

\section{Experimental bins}
\label{sec:bins}

The Belle18\,\cite{Belle:2018ezy} and Belle23\,\cite{Belle:2023bwv} experimental data are given in the form of 10-bins distribution for each of the four kinematical variables $x = \{ w, \cos \theta_l, \cos \theta_v, \chi \}$, where
\bea
       \label{eq:bins}
       \{ w_n\} & = & \{1, 1.05, 1.10, 1.15, 1.20, 1.25, 1.30, 1.35, 1.40, 1.45, w_{max} \} ~ , ~ \nonumber \\[2mm]
       \{ \mbox{cos}(\theta_v)_n \} & = & \{ -1, -0.8, -0.6, -0.4,- 0.2, 0.0, 0.2, 0.4, 0.6, 0.8, 1.0\} ~ , ~ \\[2mm]
       \{ \mbox{cos}\theta(_\ell)_n \} & = & \{ -1, -0.8, -0.6, -0.4, -0.2, 0.0, 0.2, 0.4, 0.6, 0.8, 1.0\} ~ , ~\nonumber  \\[2mm]
        \{ \chi_n\} & = & \{ 0, \frac{\pi}{5}, \frac{2\pi}{5}, \frac{3\pi}{5}, \frac{4\pi}{5}, \pi, \frac{6\pi}{5}, \frac{7\pi}{5}, \frac{8\pi}{5}, \frac{9\pi}{5}, 2\pi\} ~ . ~ \nonumber
\eea
 The BelleII23 data\,\cite{Belle-II:2023okj} are given in the same 10 bins for the variables $x = \{w,  \cos \theta_v$, $\chi \}$, while in the case of $x = \cos \theta_\ell$ the BelleII23 bins are only 8, since the first BelleII23 bin corresponds to the sum of the first three Belle18 and Belle23 bins and the BelleII23 bins $2 - 8$ correspond to the Belle18 and Belle23 bins $4 - 10$.
 Thus, we have a total of $N = 40$ data points for both Belle18 and Belle23 sets and $N = 38$ data points for BelleII23 set, including the corresponding experimental covariance matrix of dimension $N \times N$.

For the case of Belle18, using multivariate Gaussian distributions for the experimental values of $\Delta \Gamma_n^x$, we construct the ratios\,(\ref{eq:ratios}) and evaluate also the corresponding covariance matrix $\mathbf{C}_{nm}$ ($n, m = 1, ..., N$). 

Since the ratios\,(\ref{eq:ratios}) satisfy by construction the normalization condition $\sum_{n = 1}^{N_x} R_n(x) = 1$, the covariance matrix $\mathbf{C}_{nm}$ of each experiment has 4 null eigenvalues.

Using the experimental bins\,(\ref{eq:bins}) the explicit formulae for the ratios $R_n(x)$, defined in Eq.\,(\ref{eq:ratios}), read as (see also Ref.\,\cite{Martinelli:2024vde}) 
\bea
        \label{eq:ratio_w2}
        R_n(w) & = & \frac{1}{(1 + \eta)}  ~ \frac{ \widehat{H}_{++}(n)+ \widehat{H}_{--}(n) + \widehat{H}_{00}(n)}{\overline{H}_{00}} ~ , ~ \\[2mm]
        \label{eq:ratio_v2}
       R_n(\mbox{cos}\theta_v) & = & \frac{3}{20(1+ \eta) } ~ \left[ \eta+ \frac{2 - \eta}{75} \left( 91 - 33 n + 3 n^2 \right) \right] ~ , ~ \\[2mm]
         \label{eq:ratio_ell2}
        R_n(\mbox{cos}\theta_\ell) & = & \frac{3}{40(1+ \eta)} ~ \left[ 2 + \eta - \frac{\delta}{5} \left( -11 + 2 n \right) - 
                                                             \frac{2 - \eta}{75} \left( 91 - 33 n + 3 n^2 \right) \right] ~ , ~ \quad \\[2mm]
        \label{eq:ratio_chi2}
        R_n(\chi) & = & \frac{1}{10} - \frac{1}{4\pi} \frac{\epsilon}{1 + \eta} \left[ \mbox{sin}\frac{2n \pi}{5} - \mbox{sin}\frac{2(n -1) \pi}{5} \right] ~ , ~
\eea
where
\be
     \label{eq:Delta_Hii}
     \widehat{H}_{++, --, 00}(n) =  \int_{w_{n-1}}^{w_n} dw \sqrt{w^2 - 1} (1 - 2r w + r^2) ~ H_{+,-,0}^2(w) ~ . ~
\ee

\section{Asymmetries in the light-lepton sector}
\label{sec:AFB}

\begin{figure}[htb!]
\begin{center}
\includegraphics[scale=0.30]{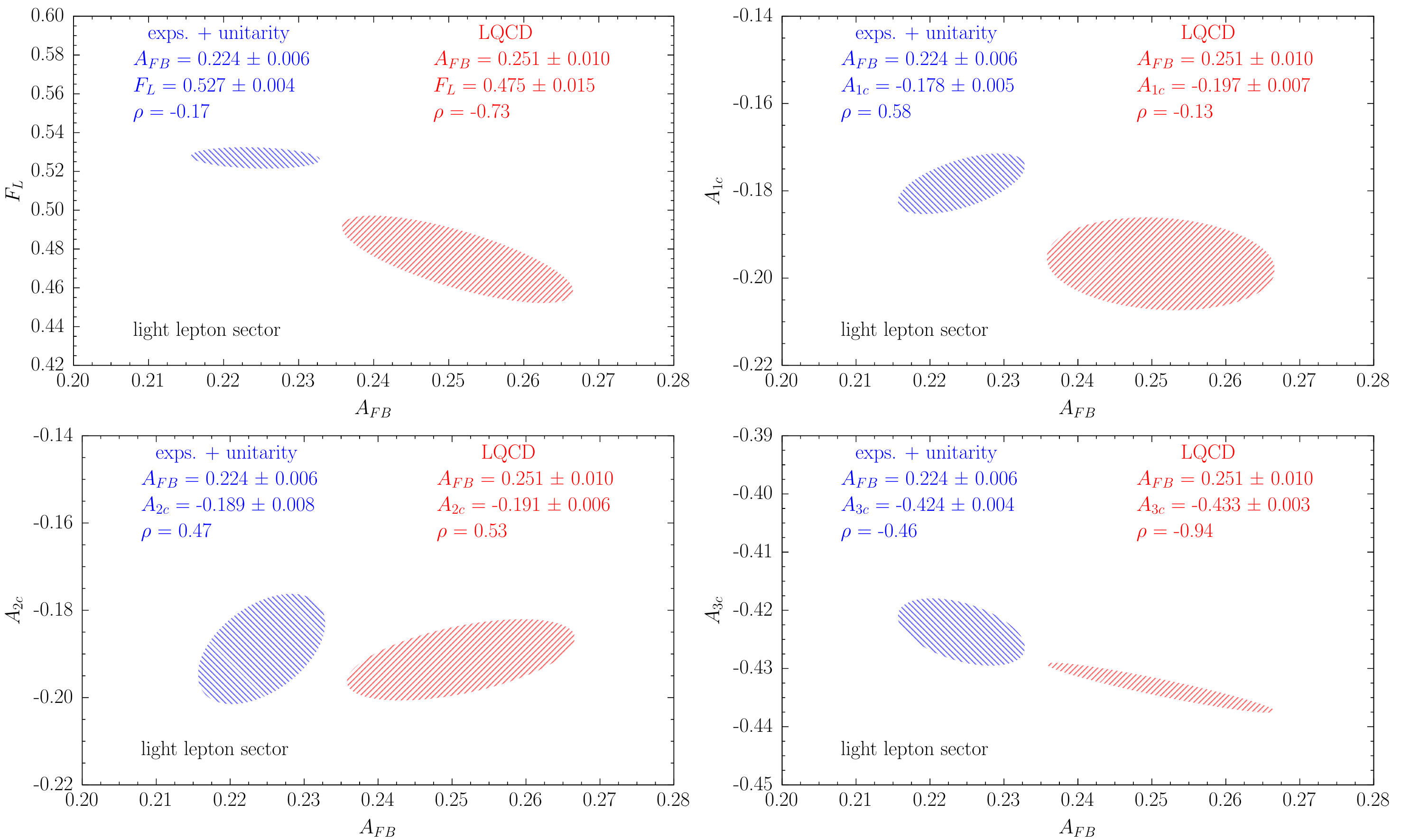}
\end{center}
\vspace{-0.5cm}
\caption{\it \small The same as in Fig.\,\ref{fig:Hij}, but for the observables $A_{FB}$, $F_L$ and $A_{1c}$ (see Eqs.(\ref{eq:AFB})-(\ref{eq:A1c})).}
\label{fig:AFB}
\end{figure}

The results for the asymmetries $A_{FB}$, $F_L$, $A_{1c}$, $A_{2c}$ and $A_{3c}$, given in Eqs.\,(\ref{eq:AFB})-(\ref{eq:A3c}), are collected in Table\,\ref{tab:AFB} and visualized as contour plots in Fig.\,\ref{fig:AFB}, including the correlations among the various quantities. 
For completeness, we collect in Table\,\ref{tab:AFB} also the results obtained for the three hadronic parameters $\eta$, $\delta$ and $\epsilon$, given in Eqs.\,(\ref{eq:eta})-(\ref{eq:epsilon}). 

All the results compare positively (i.e., differences less than one standard deviation) with the corresponding results from Ref.\,\cite{Martinelli:2024vde}, where the SM and the massless lepton limit are not assumed.
In particular, the differences between the three asymmetries $A_{FB}$, $F_L$ and $A_{1c}$, obtained using simultaneously all the three experimental data sets, and those corresponding to the LQCD predictions are above the $2 \sigma$ level and may reach $\simeq 3.3 \sigma$ in the case of $F_L$. On the contrary, they do not exceed $\simeq 1.4 \sigma$ in the case of the individual Belle23 data set.

\begin{table}[htb!]
\renewcommand{\arraystretch}{1.5}
\begin{center}
\begin{adjustbox}{max width=\textwidth}
\begin{tabular}{|c||c|c|c|c|c||c|c|c||}
\hline
& $A_{FB}$ & $F_L$ & $A_{1c}$ & $A_{2c}$ & $A_{3c}$ & $\eta$ & $\delta$ & $\epsilon$ \\ \hline
\hline
Belle18                                                & 0.218 (10) & 0.530 ~(7) & -0.177 ~(9) & -0.198 (16) & -0.423 ~(9) & 0.885 (24) & -0.549 (28) & 0.335 (18) \\ \hline
Belle23                                                & 0.231 (13) & 0.502 (12) & -0.187 (13) & -0.197 (19) & -0.409 (11)  & 0.995 (50) & -0.615 (37) & 0.373 (32) \\ \hline
BelleII23                                              & 0.200 (10) & 0.528 ~(5) & -0.188 ~(7) & -0.202 (11) & -0.417 ~(5) & 0.895 (17) & -0.507 (26) & 0.356 (14) \\ \hline \hline
Belle18 + Belle23 + BelleII23              & 0.224 ~(6) & 0.527 ~(4) & -0.178 (~5) & -0.189 ~(8) & -0.424 (~4) & 0.898 (13) & -0.567 (15) & 0.339 (10) \\ \hline \hline
LQCD                                                  & 0.251 (10) & 0.475 (15) & -0.197 ~(7) & -0.191 ~(6) & -0.433 ~(3) & 1.109 (66) & -0.707 (48) & 0.415 (26) \\ \hline
\end{tabular}
\end{adjustbox}
\end{center}
\renewcommand{\arraystretch}{1.0}
\vspace{-0.25cm}
\caption{\it \small The same as in Table\,\ref{tab:Hij}, but for the (light-lepton) asymmetries $A_{FB}$, $F_L$, $A_{1c}$, $A_{2c}$ and $A_{3c}$, given in Eqs.\,(\ref{eq:AFB})-(\ref{eq:A3c}), and for the three hadronic parameters $\eta$, $\delta$ and $\epsilon$, given in Eqs.\,(\ref{eq:eta})-(\ref{eq:epsilon}). The LQCD results for $A_{FB}$ and $F_L$ correspond to the ones labelled ``Combined" in Table 4 of Ref.\,\cite{Martinelli:2023fwm}.}
\label{tab:AFB}
\end{table}

\section{Differential ratios $(1 / \Gamma) d\Gamma / dx$ for the $\tau$-lepton}
\label{sec:ratios_tau}

In this Appendix, while keeping the SM framework, we provide the generalization of Eqs.\,(\ref{eq:ratio_w})-(\ref{eq:ratio_chi}) to the case of the massive $\tau$-lepton.
For $x = w$ one has
\bea
        \label{eq:ratio_w_tau}
        \frac{1}{\Gamma} \frac{d\Gamma}{dw} & = & \sqrt{w^2 - 1} \left( 1 - \frac{m_\tau^2}{q^2} \right)^2 \frac{1}{\overline{\cal{H}}} \left\{ \frac{3m_\tau^2}{2m_B^2} H_t^2(w) \right. \nonumber \\[2mm]
                                                                      & + & \left. \left( 1 - 2 r w + r^2 + \frac{m_\tau^2}{2m_B^2} \right) \left[ H_+^2(w) + H_-^2(w) + H_0^2(w) \right] \right \} ~ , ~
\eea
where $\overline{\cal{H}}$ is defined in Eq.\,(\ref{eq:H}).
In the case of the angular variables $x = \{ \cos \theta_l, \cos \theta_v, \chi \}$ we follow Ref.\,\cite{Martinelli:2024vde} obtaining
\bea
      \label{eq:thetav_tau}
      \frac{1}{\Gamma} \frac{d\Gamma}{d\mbox{cos}\theta_v} & = &  \frac{3}{4 (1 + \eta_\tau)} \left\{ \eta_\tau + (2 - \eta_\tau) \mbox{cos}^2\theta_v \right\} ~ , ~ \\[2mm]
      \label{eq:thetal_tau}
       \frac{1}{\Gamma} \frac{d\Gamma}{d\mbox{cos}\theta_\ell} & = &  \frac{3}{8 (1 + \eta_\tau^\prime)} \left\{ 2 + \eta_\tau^\prime -2 \delta_\tau \mbox{cos}\theta_\ell - 
                  (2 - \eta_\tau^\prime) \mbox{cos}^2\theta_\ell \right\} ~ , ~ \\[2mm]
      \label{eq:chi_tau}
       \frac{1}{\Gamma} \frac{d\Gamma}{d\chi} & = & \frac{1}{2 \pi} \left\{ 1 - \frac{\epsilon_\tau}{1 + \eta_\tau} \mbox{cos}2\chi \right\} ~ , ~
\eea 
where the quantities $\{ \eta_\tau, \eta_\tau^\prime, \delta_\tau, \epsilon_\tau \}$ are explicitly given by
\bea
     \label{eq:eta_SM}
     \eta_\tau & = & \frac{\overline{H}_{++}^\tau + \overline{H}_{--}^\tau + \frac{m_\tau^2}{2 m_B^2} \left( \overline{L}_{++}^\tau + \overline{L}_{--}^\tau \right)}
                              {\overline{H}_{00}^\tau +  \frac{m_\tau^2}{2 m_B^2} \left( \overline{L}_{00}^\tau + 3  \overline{L}_{tt}^\tau \right)} ~ , ~ \\[2mm]
     \label{eq:etap_SM}
     \eta_\tau^\prime & = & \frac{\overline{H}_{++}^\tau + \overline{H}_{--}^\tau + \frac{m_\tau^2}{m_B^2} \left( \overline{L}_{00}^\tau + \overline{L}_{tt}^\tau \right)}
                      {\overline{H}_{00}^\tau +  \frac{m_\tau^2}{2 m_B^2} \left( \overline{L}_{++}^\tau + \overline{L}_{--}^\tau -\overline{L}_{00}^\tau + \overline{L}_{tt}^\tau \right)} ~ , ~ \\[2mm]
     \label{eq:delta_SM}
     \delta_\tau & = & \frac{\overline{H}_{++}^\tau - \overline{H}_{--}^\tau + \frac{2m_\tau^2}{m_B^2} \overline{L}_{0t}^\tau}
                                {\overline{H}_{00}^\tau +  \frac{m_\tau^2}{2 m_B^2} \left( \overline{L}_{++}^\tau + \overline{L}_{--}^\tau -\overline{L}_{00}^\tau + \overline{L}_{tt}^\tau \right)} ~ , ~ \\[2mm]
     \label{eq:epsilon_SM}
     \epsilon_\tau & = &  \frac{\overline{H}_{+-}^\tau - \frac{m_\tau^2}{m_B^2} \overline{L}_{+-}^\tau}
                                     {H_{00} +  \frac{m_\tau^2}{2 m_B^2} \left( \overline{L}_{00}^\tau + 3  \overline{L}_{tt}^\tau \right)} ~ , ~  
\eea
with $\overline{H}_{ij}^\tau$ and $\overline{L}_{ij}^\tau$ being defined in Eqs.\,(\ref{eq:Hii_tau})-(\ref{eq:Lij_tau}) for $i = j = +, -, 0, t$ (and straightforwardly generalized to the case $i \neq j$). Note that the two quantities $\eta_\tau$ and $\eta_\tau^\prime$ differ only by lepton mass effects, namely $\eta_\tau - \eta_\tau^\prime = {\cal{O}}(m_\tau^2)$.

In terms of the quantities $\{ \eta_\tau, \eta_\tau^\prime, \delta_\tau, \epsilon_\tau \}$ the $\tau$-observables $A_{FB, \tau}$, $F_{L, \tau}$ and $A_{1c, \tau}$ considered in this work can be written as
\bea
       \label{eq:AFB_tau}
       A_{FB, \tau} & = & -\frac{3}{4} \frac{\delta_\tau}{1 + \eta_\tau^\prime} ~ , ~ \\[2mm]
       \label{eq:FL_tau}
       F_{L, \tau} & = & \frac{1}{1 + \eta_\tau} ~ , ~ \\[2mm]
       \label{eq:A1c_tau}
       A_{1c, \tau} & = & - \frac{\epsilon_\tau}{1 + \eta_\tau} ~ , ~
\eea
while $R(D^*)$ and $P_\tau(D^*)$ are given by Eqs.\,(\ref{eq:RDstar}) and \,(\ref{eq:PDstar_tau}), respectively.

\section{Correlation matrices}
\label{sec:corr}

In order to use our results for further phenomenological analyses we provide in Table\,\ref{tab:correxp} the mean values with uncertainties and the correlation matrix among our predictions for the observables $A_{FB, \ell}$, $F_{L, \ell}$, $A_{1c, \ell}$, $R(D^*)$, $F_{L, \tau}$, $(F_{L, \tau})_{<,>}$, $P_\tau(D^*)$ and $A_{FB, \tau}$, obtained in Section\,\ref{sec:light+tau} by our unitary BGL fit of the light-lepton experimental ratios $R_n(x)$\,\cite{Belle:2018ezy, Belle:2023bwv, Belle-II:2023okj} plus the experimental values of $R(D^*)$\,\cite{Banerjee:2024znd} and $(F_{L, \tau})_{<,>}$\,\cite{LHCb:2023ssl}.

\begin{table}[htb!]
\renewcommand{\arraystretch}{1.5}
\begin{center}
\begin{adjustbox}{max width=\textwidth}
\begin{tabular}{|c|c||c|c|c||c|c|c|c|c|c||}
\hline
& & $A_{FB, \ell}$ & $F_{L, \ell}$ & $A_{1c, \ell}$ & $R(D^*)$ & $F_{L, \tau}$ & $(F_{L, \tau})_>$ &  $(F_{L, \tau})_<$ & $P_\tau(D^*)$ & $A_{FB, \tau}$ \\ \hline
\hline
$A_{FB, \ell}$              &  0.2244 (57) &   1.0000 &  -0.1684 &   0.5906  &  0.0044 & -0.0415  & -0.0451 & -0.0356  & -0.0150 &  0.2271  \\ \hline
$F_{L, \ell}$                 &  0.5266 (36) &  -0.1684 &  1.0000  &  0.3272  & -0.0165  &  0.1136  &  0.1011 &   0.1350  &  0.0032 & -0.0014   \\ \hline
$A_{1c, \ell}$               & -0.1786 (46) &   0.5906 &  0.3272  &  1.0000  &  0.0087  &  0.0861  &  0.1020 &   0.0834  & -0.0034 & -0.1051  \\ \hline \hline
$R(D^*)$                     &  0.277 (11)    &   0.0044 & -0.0165 &   0.0087  &  1.0000  &  0.9688 &  0.9531  &   0.9755  &  0.9897 & -0.8501   \\ \hline
$F_{L, \tau}$               &  0.500 (20)    & -0.0415  &  0.1136  &  0.0861  &  0.9688  &  1.0000 &  0.9957  &   0.9953  &  0.9751 & -0.8890 \\ \hline
$(F_{L, \tau})_>$         &  0.438 (18)    & -0.0451 &  0.1011  &  0.1020  &  0.9531  &  0.9957  &  1.0000  &   0.9830  &  0.9562 & -0.8976   \\ \hline
$(F_{L, \tau})_<$         &  0.600 (23)    & -0.0356 &  0.1350  &  0.0834  &  0.9755  &  0.9953  &  0.9830  &   1.0000  &  0.9807 & -0.8729   \\ \hline
$P_\tau(D^*)$              & -0.372 (55)   & -0.0150 &  0.0032  & -0.0034  &  0.9897  &  0.9751  &  0.9562  &   0.9807  &  1.0000 & -0.8589   \\ \hline
$A_{FB, \tau}$             &  0.003 (26)   &  0.2271 &-0.0014   &--0.1051  & -0.8501  & -0.8890  & -0.8976  & -0.8729   & -0.8589 & 1.0000  \\ \hline
\end{tabular}
\end{adjustbox}
\end{center}
\renewcommand{\arraystretch}{1.0}
\vspace{-0.25cm}
\caption{\it \small The mean values with uncertainties and the correlation matrix among our predictions for the observables $A_{FB, \ell}$, $F_{L, \ell}$, $A_{1c, \ell}$, $R(D^*)$, $F_{L, \tau}$, $(F_{L, \tau})_{<,>}$, $P_\tau(D^*)$ and $A_{FB, \tau}$, obtained in Section\,\ref{sec:light+tau} by our unitary BGL fit of the light-lepton experimental ratios $R_n(x)$\,\cite{Belle:2018ezy, Belle:2023bwv, Belle-II:2023okj} plus the experimental values of $R(D^*)$\,\cite{Banerjee:2024znd} and $(F_{L, \tau})_{<,>}$\,\cite{LHCb:2023ssl}.}
\label{tab:correxp}
\end{table}

In Table\,\ref{tab:corrth} we collect our results in the case of the theoretical predictions of the unitary DM method of Ref.\,\cite{Martinelli:2023fwm} applied to all available LQCD simulations of the hadronic FFs from FNAL/MILC\,\cite{FermilabLattice:2021cdg}, HPQCD\,\cite{Harrison:2023dzh} and JLQCD\,\cite{Aoki:2023qpa} Collaborations.

\begin{table}[htb!]
\renewcommand{\arraystretch}{1.5}
\begin{center}
\begin{adjustbox}{max width=\textwidth}
\begin{tabular}{|c|c||c|c|c||c|c|c|c|c|c||}
\hline
& & $A_{FB, \ell}$ & $F_{L, \ell}$ & $A_{1c, \ell}$ & $R(D^*)$ & $F_{L, \tau}$ & $(F_{L, \tau})_>$ &  $(F_{L, \tau})_<$ & $P_\tau(D^*)$ & $A_{FB, \tau}$  \\ \hline
\hline
$A_{FB, \ell}$              &  0.2512 (102) &  1.0000 & -0.7346  & -0.1298  & 0.4214  & -0.7205 & -0.6951 & -0.7367 & -0.5216  &  0.9692 \\ \hline
$F_{L, \ell}$                 &  0.4747 (149) & -0.7346 &  1.0000  & 0.7667   & -0.7387 &  0.9722 &  0.9129 &  0.9818  &  0.8807  & -0.7761 \\ \hline
$A_{1c, \ell}$               & -0.1967 (70)   & -0.1298 &  0.7667  & 1.0000   & -0.6611 &  0.7408 &  0.6791 &  0.7419  &  0.7853  & -0.2243  \\ \hline\hline
$R(D^*)$                     &  0.2581 (48)    &  0.4214 & -0.7387  &-0.6611   & 1.0000  & -0.6548 & -0.5678 & -0.6240 & -0.7758  &  0.4450  \\ \hline
$F_{L, \tau}$               &  0.4257 (77)    & -0.7205 &   0.9722  & 0.7408  & -0.6548 &  1.0000 &  0.9816 &  0.9922  &  0.8868  & -0.7989  \\ \hline
$(F_{L, \tau})_>$         &  0.3843 (45)    & -0.6951 &   0.9129 &  0.6791 & -0.5678  &  0.9816 &  1.0000 &  0.9584  &  0.8332  & -0.7922  \\ \hline
$(F_{L, \tau})_<$         &  0.4992 (125)  & -0.7367 &   0.9818  & 0.7419 & -0.6240  &  0.9922 &  0.9584 &  1.0000  &  0.8736  & -0.8016  \\ \hline
$P_\tau(D^*)$              & -0.5208 (59)    & -0.5216 &  0.8807  & 0.7853   & -0.7758 &  0.8868 &  0.8332 &  0.8736 &  1.0000   & -0.6319  \\ \hline
$A_{FB, \tau}$             &  0.0781 (79)    &  0.9692 &  -0.7761 & -0.2243   & 0.4450 & -0.7989 & -0.7922 & -0.8016 & -0.6319   &  1.0000  \\ \hline
\end{tabular}
\end{adjustbox}
\end{center}
\renewcommand{\arraystretch}{1.0}
\vspace{-0.25cm}
\caption{\it \small The same as in Table\,\ref{tab:correxp}, but in the case of the theoretical predictions of the unitary DM method of Ref.\,\cite{Martinelli:2023fwm} applied to all available LQCD simulations of the hadronic FFs from FNAL/MILC\,\cite{FermilabLattice:2021cdg}, HPQCD\,\cite{Harrison:2023dzh} and JLQCD\,\cite{Aoki:2023qpa} Collaborations.}
\label{tab:corrth}
\end{table}

\section{Ratios of HQET-inspired form factors}
\label{sec:Ri}
In the literature it is common to consider the following ratios among the HQET-inspired FFs\,(\ref{eq:hA1}):
\bea
     \label{eq:R0}
     R_0(w) & = & \frac{1}{1+r} \left[ 1 + w + (r w -1) \frac{h_{A_2}(w)}{h_{A_1}(w)} + (r - w) \frac{h_{A_3}(w)}{h_{A_1}(w)} \right]  ~ , ~ \nonumber \\[2mm]
     \label{eq:R1}
     R_1(w) & = & \frac{h_V(w)}{h_{A_1}(w)} ~ , ~ \\[2mm]
     \label{eq:R2}
     R_2(w) & = & \frac{r h_{A_2}(w) + h_{A_3}(w)}{h_{A_1}(w)} ~ . ~ \nonumber
\eea
All the above ratios become equal to unity in the limit of infinite heavy-quark masses.
 
\begin{figure}[htb!]
\begin{center}
\includegraphics[scale=0.30]{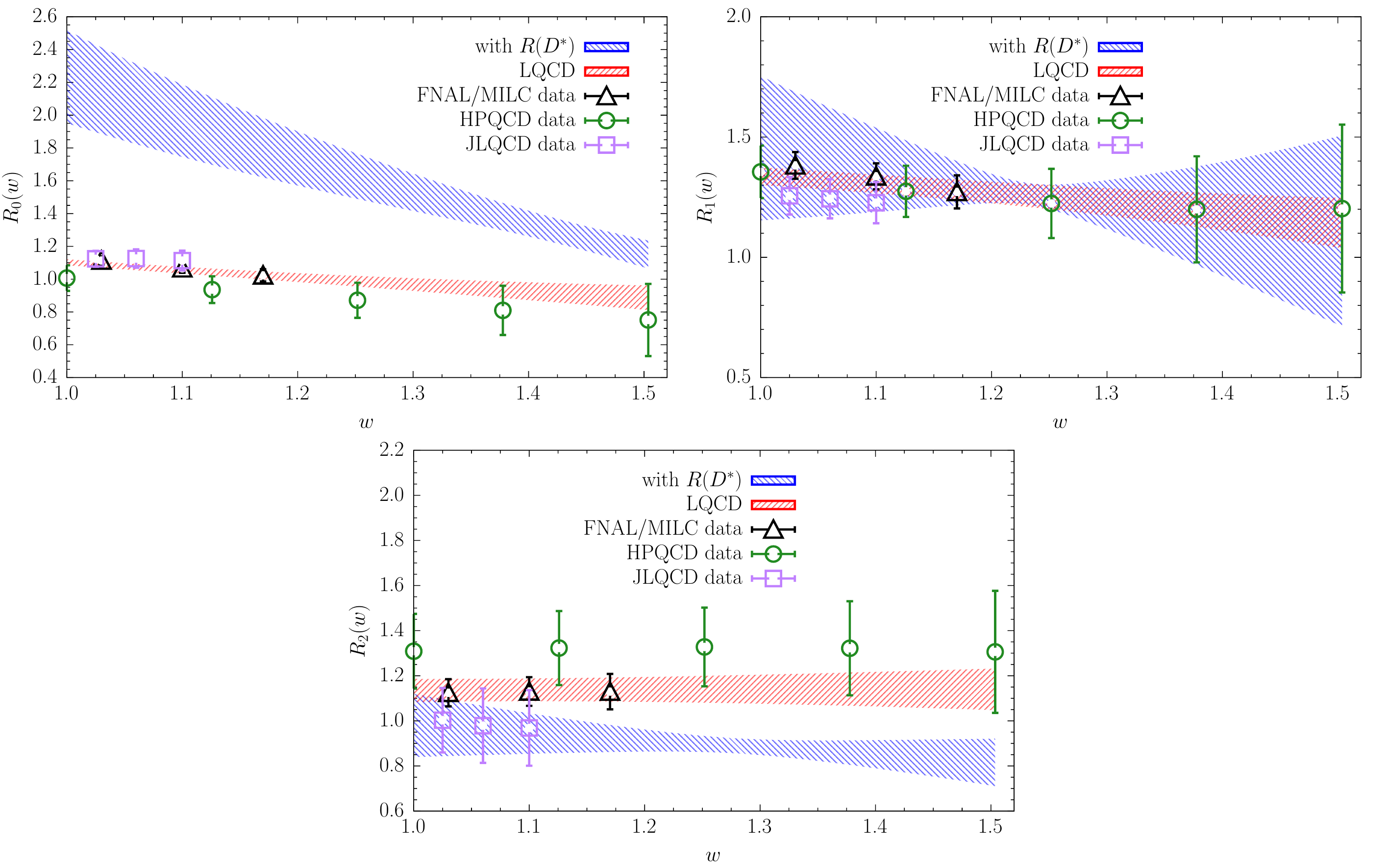}
\end{center}
\vspace{-0.5cm}
\caption{\it \small Comparison of the ratios $R_0(w)$, $R_1(w)$ and $R_2(w)$ , defined by Eqs.\,(\ref{eq:R1}), corresponding to the reduced FFs extracted within the SM from the (light-lepton) experimental ratios $R_n(x)$ plus the experimental value of $R(D^*)$, with the individual LQCD results from FNAL/MILC\,\cite{FermilabLattice:2021cdg}, HPQCD\,\cite{Harrison:2023dzh} and JLQCD\,\cite{Aoki:2023qpa} Collaborations, as well as with the LQCD predictions, obtained by the unitary DM method of Ref.\,\cite{Martinelli:2023fwm} applied to all available LQCD data.}
\label{fig:Ri}
\end{figure}

In Fig.\,\ref{fig:Ri} we show the results for the ratios $R_0(w)$, $R_1(w)$ and $R_2(w)$ corresponding to the FFs extracted within the SM from the (light-lepton) experimental ratios $R_n(x)$ plus the experimental value of $R(D^*)$. The comparison with the individual LQCD results from FNAL/MILC\,\cite{FermilabLattice:2021cdg}, HPQCD\,\cite{Harrison:2023dzh} and JLQCD\,\cite{Aoki:2023qpa} Collaborations, as well as with the LQCD predictions, obtained by the unitary DM method of Ref.\,\cite{Martinelli:2023fwm} applied to all available LQCD data, indicates that the experimental value of $R(D^*)$ would require within the SM huge deviations from the heavy-quark limit for the ratio $R_0(w)$.
Note also the significant spread of values for $R_2(w)$ among different LQCD Collaborations.
Similar results have been shown recently in Ref.\,\cite{Jung24}.

\bibliography{biblio}
\bibliographystyle{EPJC}

\end{document}